\documentclass[sn-basic]{sn-jnl}

\usepackage{graphicx}%
\usepackage{multirow}%
\usepackage{amsmath,amssymb,amsfonts}%
\usepackage{amsthm}%
\usepackage{mathrsfs}%
\usepackage[title]{appendix}%
\usepackage{xcolor}%
\usepackage{textcomp}%
\usepackage{manyfoot}%
\usepackage{booktabs}%
\usepackage{algorithm}%
\usepackage{algorithmicx}%
\usepackage{algpseudocode}%
\usepackage{listings}%
\usepackage{mathptmx}
\usepackage{bm}
\newcommand{\DIVm}{\bm{\nabla} \cdot }
\newcommand{\pd}{\partial}
\newcommand{\ut}{{\bm u}}
\newcommand{\uf}{{\bm u'}}
\newcommand{\um}{\overline{{\bm u}}}
\newcommand{\Bt}{{\bm B}}
\newcommand{\Bf}{{\bm b'}}
\newcommand{\Bm}{\overline{{\bm B}}}
\newcommand{\Bmj}{\overline{B}_j}
\newcommand{\emf}{\mbox{\boldmath ${\cal E}$} {}}
\newcommand{\emfi}{\mbox{${\cal E}_i$} {}}
\newcommand{\ddelta}{\mbox{\boldmath $\delta$} {}}
\newcommand{\ggamma}{\mbox{\boldmath $\gamma$} {}}

\newcommand{\kkappa}{\mbox{\boldmath $\kappa$} {}}

\newcommand{\nab}{\mbox{\boldmath $\nabla$} {}}

\newcommand\kpc{~ {\rm kpc}}

\newcommand\OSN{~ \Omega_{\rm Sn} }
\newcommand\ZSN{~ \sigma_{\rm Sn} }
\newcommand\Pm{{\rm Pr}_{\rm M}}
\newcommand\Rm{{\rm Re}_{\rm M}}
\renewcommand\Re{\rm Re}

\newcommand{\mjk}[1]{\textcolor{brown}{#1}}

\begin{document}

\title[Dynamos in galaxies]{Computational approaches to modeling dynamos in galaxies}

\author*[1,3,4]{\fnm{Maarit J.} \sur{Korpi-Lagg}}\email{maarit.korpi-lagg@aalto.fi}
\author[2]{\fnm{Mordecai-Mark} \sur{Mac Low}}\email{mordecai@amnh.org}
\author[1,3,5]{\fnm{Frederick A.} \sur{Gent}}\email{frederick.gent@aalto.fi}

\affil[1]{\orgdiv{Astroinformatics, Department of Computer Science}, \orgname{Aalto University}, \orgaddress{\street{P.O. Box 15400}, \postcode{FI-00076}, \city{Espoo}, \country{Finland}}}
\affil[2]{\orgdiv{Department of Astrophysics}, \orgname{American Museum of Natural History}, \orgaddress{\street{200 Central Park West}, \city{New York}, \postcode{NY 10024}, \country{USA}}}
\affil[3]{\orgname{Nordic Institute for Theoretical Physics}, \orgaddress{\street{Hannes Alfv\'ens väg 12}, \postcode{SE-106 91} \city{Stockholm}, \country{Sweden}}}
\affil[4]{\orgname{Max Planck Institute for Solar System Research}, \orgaddress{\street{Justus-von-Liebig-Weg 3}, \postcode{D-37707} \city{G\"ottingen}, \country{Germany}}}
\affil[5]{\orgdiv{School of Mathematics, Statistics and Physics}, \orgname{Newcastle University}, \orgaddress{\postcode{NE1 7RU}, \city{Newcastle upon Tyne}, \country{UK}}}

\abstract{
Galaxies are observed to host magnetic fields with a typical total strength of
around 15$\mu$G. A coherent large-scale field constitutes up to a few
microgauss of the total, while the rest is built from strong magnetic
fluctuations over a wide range of spatial scales.  This represents sufficient
magnetic energy for it to be dynamically significant. Several questions
immediately arise: What is the physical mechanism that gives rise to such
magnetic fields? How do these magnetic fields affect the formation and
evolution of galaxies? In which physical processes do magnetic fields play a
role, and how can that role be characterized?  Numerical modelling of
magnetized flows in galaxies is playing an ever-increasing role in finding
those answers. We review major techniques used for these models.
Current results strongly support the conclusion that field
growth occurs during the formation of the first galaxies on timescales shorter
than their accretion timescales due to small-scale turbulent dynamos.  The
saturated small-scale dynamo maintains field strengths at a few percent of
equipartition with turbulence.  The subsequent action of large-scale dynamos in
differentially rotating discs produces observed modern field strengths in
equipartition with the turbulence and having power at large
scales. The field structure resulting appears consistent with
observations including Faraday rotation and polarisation from
synchrotron and dust thermal emission. Major remaining challenges
include scaling numerical models toward realistic scale separations and Prandtl and Reynolds
numbers.}


\maketitle

\setcounter{tocdepth}{3}
\tableofcontents

\section{Introduction}

Galaxies are observed to host magnetic fields with a typical total strength of
around 15$\mu$G. A coherent large-scale field constitutes up to a few
microgauss of the total, while the rest is built from strong magnetic
fluctuations over a wide range of spatial scales \citep{Beck15,beck2019}.
Until recently, the importance of magnetic fields has often been overlooked
when considering the dynamics and evolution of galaxies.  However, magnetic
field energy is sufficiently large to have a significance comparable to
kinetic energy.  Several questions immediately arise. What are the physical
mechanisms that give rise to such strong magnetic fields?
How do different physical processes in galaxies contribute to magnetic
field growth?  What are the small and large-scale structures of these fields?
It is becoming increasingly clear that both small-scale dynamo
  (SSD) and large-scale dynamo (LSD) processes play a fundamental role
in determining the answers to these questions.  Progress in
understanding dynamo behavior relies ever more strongly on numerical
modelling of the magnetised plasma in galaxies.
This review summarises the
current state-of-the-art of these numerical simulations and their results.  See
also the recent review by \cite{brandenburg2022}.

\subsection{Fundamental observations of galactic magnetic fields} \label{sect:obs}

In galaxies, both the magnetic field and the velocity have structure
across a wide spectrum of characteristic scales.  First, there are
contributions from the largest scales, such as the differential
angular velocity that arises from the typical flat rotation curves
with nearly constant rotation velocity as a function of radius
observed in spiral galaxies embedded in dark matter halos.  At
slightly smaller scales, galactic outflows from galaxy centers or
from powerful superbubbles produced by correlated SN explosions from
young star clusters produce ordered velocities in the vertical
direction outward from the galactic disc plane.  Similarly, the
magnetic field of spiral galaxies has been verified to have in most
cases a coherent large-scale structure that sometimes, but by no
means always, follows the optical spiral arms formed by the stars.

We have a variety of indirect
observations through which we can identify the properties of the magnetic
field in the interstellar medium (ISM), circumgalactic medium, intracluster medium (ICM) and, more
locally, in molecular clouds, within and between spiral arms and
around supernova (SN) remnants or superbubbles \citep[see, e.g.,
Table~1 of][]{Beck15}.  The total magnetic field
$B_{\rm tot_\perp}$ perpendicular to the line of sight is
measured through total synchrotron intensity. This is the sum of the
regular field $B_{\rm reg_\perp}$, which is defined as
consistent within the beam width of the telescope, and the locally
turbulent field $B_{\rm turb_\perp}$, which are the fluctuations
on top of the regular field within the width of the beam,
whether due to SSD, tangling or both. The regular field is
discriminated from the total field through polarised synchrotron
emission. The sum of isotropic turbulence $B_{\rm iso_\perp}$, with
dispersions matching in all three dimensions, and anisotropic
turbulence $B_{\rm aniso_\perp}$, where dispersions impacted by
compressions and shear differ, comprises $B_{\rm turb_\perp}$.

The isotropic turbulent field can be distinguished by combining
unpolarised synchrotron intensity, beam depolarisation, and Faraday
depolarisation \citep[see, e.g.,][]{Ferriere20}.  The anistropic
turbulent field can be difficult to distinguish from the regular
field, so the ordered field energy $B_{\rm ord_\perp}^2=B_{\rm
  reg_\perp}^2+B_{\rm aniso_\perp}^2$.  This is determined via
intensity and radio, optical, infrared (IR), and submillimeter
polarisation.  The Goldreich-Kylafis \citep{GK81} and longitudinal
Zeeman effects provide $B_{\rm reg_\perp}$ and $B_{\rm
  reg_\parallel}$, the latter being parallel to the line of sight.

When interpreting observations, there is an underlying assumption that
total magnetic field and total cosmic ray energy are in equipartition
in regions emitting synchrotron radiation.  There are various
complications to deriving the field strength from the synchrotron
intensity, which include the increased energy losses in cosmic rays
(CRs), particularly electrons, in higher density environments, such as
starburst regions and massive spiral arms, and the differences in the
energy distribution of CRs produced by various sources, from which the
integrated CR energy is obtained.  Notwithstanding such uncertainties,
a survey of 21 nearby galaxies obtained $B_{\rm tot}=17\pm13\,\mu$G,
including $B_{\rm reg} \simeq=5\pm3\,\mu$G \citep{Fletcher10a}.
Fields as strong as 50--100 $\mu$G are observed in starburst galaxies
and some barred galaxies \citep{Beck05,HBKD11,AKKWBD13}.  Such field
strengths clearly indicate a magnetic field capable of exerting
significant influences on the evolving dynamics and structure of the
galaxies, including the processes leading to star formation. 
The relatively high strength of the turbulent field
relative to the regular field is a common feature across observations
on the scale of the Galaxy \cite[see, e.g., Table~3][]{beck2019}, who
determine $0.04\leq B_{\rm reg}/B_{\rm tot}\leq 0.43$ across 12
galaxies and at a variety of galactic radii.

Even setting aside the uncertainties in the measurements of the
magnetic field, there is no observational diagnostic beyond its
overall magnitude to distinguish the turbulent field that arises
through an SSD from that which results from tangling of the regular
field by the turbulent flow.  Recent high resolution observations of
extragalactic magnetic fields with {\em SOFIA} \citep{SOFIA21}
identify polarisation fractions in the dense star forming regions
traced by far-infrared (FIR) anticorrelated with the star formation
rate, and generally lower than the polarisation of 3 and 6 cm radio
emission.  This is suggested to imply complex magnetised structures in
the star-forming regions characteristic of SSDs acting there.  An
accurate model of the turbulent structure of the magnetic field in
galaxies is required, not only to measure and understand the magnetic
field, but also to correctly interpret its effect on how observables
from distant sources reach the observer.

\cite{TEEMP23} use a 3D map of observed dust density distribution in the
local bubble of the Solar neighbourhood \citep{LE19,LGE20} and simulated cosmic
ray detection to infer the structure of the local magnetic field, assuming weak
turbulence. Such observational inference provides a suitable benchmark against
which to test numerical models.  Insight derived from numerical modelling into
the correlation between the field and velocities, gas and dust distributions
will improve observational inference. From extragalactic sources
\cite{Brown07} use RM to determine the mean magnetic field in the Milky Way.
It aligns along spiral arms and has a strength up to about $2\mu$G.
Within alternating spiral arms it is observed to switch sign.

Such field reversals are not common in observations of external galaxies.
Determinations of the magnetic field in M31 \citep{Fletcher:2004} and M51
\citep{Fletcher:2011} show the mean and fluctuating fields to be of roughly
equivalent energy, with strength $5\mu$G each across a radial extent between 6
and 14 kpc.
Non-axisymmetric configurations are certainly observed, including
  misalignment along the spiral arms
  \citep{Fletcher:2004,Fletcher10a,Fletcher:2011} and field reversals
  \citep{Brown07,Haverkorn15}.
Pitch angles in M31 steepen from $-17\deg$ at $6 < r<10$ kpc to
$-8\deg$ at $12 < r<14$~kpc.  At high redshift, galaxies have already been
observed to have magnetic fields of order $1\mu$G \citep{bernet2008}.

\subsection{Governing physical equations} \label{sect:physics}

Galactic dynamos are most often studied using the magnetohydrodynamics
(MHD) framework, which describes
the interactions of magnetic fields with a plasma that is treated as a continuum. Further, the plasma
is usually assumed to be resistive, that is, its magnetic diffusivity
$\eta=\left(\mu_0 \sigma \right)^{-1}\neq0$,
where $\mu_0$ is the vacuum permeability and $\sigma$ the electrical
conductivity of the medium.
The justification for these approximations rests on the tiny Larmor
radius of charged particles in the ISM \begin{equation}
  r_{\rm L} = \frac{mcv}{qB} = (36 \mathrm{ km}) \left(\frac{m}{m_{\rm
      H}}\right) \left(\frac{v}{10 \mathrm{ km s}^{-1}}\right) \left(\frac{e}{q}\right)\left(\frac{3\, \mu\mathrm{G}}{B}\right)
\end{equation}
where $m$ and $q$ are
the mass and charge
of the charged particles, $v$ and $B$ are the magnitudes of
their perpendicular velocity and the ambient magnetic field,
$m_{\rm H} = 1.67 \times 10^{-24} \mathrm{ g}$ and
$e = 4.8 \times 10^{-10} \mathrm{ esu}$ are the mass and charge of a
proton, and $c = 2.99 \times 10^{10}\mathrm{ cm s}^{-1}$ is the speed of
light.  The details of reconnection have been argued to depend on kinetic effects
\citep{YKJ10}, but MHD turbulence may inevitably
drive small-scale reconnecting structures regardless of kinetic scale
effects \citep{eyink2011,oishi2015, Laz20}.
For modelling dynamo behavior the turbulent reconnection
process can be usefully abstracted with Laplacian Ohmic
resistivity.

The evolution of magnetic fields in an electrically conducting fluid is governed by the induction equation
 \begin{equation}
\frac{\partial {\bm B}}{\partial t}=\nabla \times \left({\bm u} \times {\bm B} - \nabla \times \eta \nabla \times {\bm B} \right),
\label{eq:ind}
\end{equation}
where ${\bm B}$ is the magnetic field and ${\bm u}$ is the velocity
field.
The governing equation for the dynamics of viscous fluids is the
Navier--Stokes equation, which can be written in the following
non-conservative form for rotating, magnetised galaxies in a
gravitational potential:
\begin{equation}
  \label{eq:mom}
  \rho \frac{D {\bm u}}{Dt} = \rho {\bm g} - \bm{\nabla}p - 2 \bm{\Omega} \times \bm{u} + \bm{J} \times \bm {B} + \bm{\nabla} \cdot \left( 2 \rho \nu {\sf S} \right),
\end{equation}
where ${\bm{g}}$ is the gravitational acceleration field, $p$ is the thermal
pressure, $\bm{\Omega}$ is the rotation vector, $\bm{J}$ is the
current density, $\nu$ is the kinematic viscosity, and ${\sf S}$ is
the traceless rate of strain tensor, ${\sf S}_{ij}= \frac{1}{2}
\left(u_{i,j} + u_{j,i} \right) -\frac{1}{3} \delta_{ij} u_{k,k}$.
Additional terms such as the cosmic ray pressure gradient, coupled to a model
of cosmic ray production, destruction, advection, and energy loss and
gain can also play a role.
When the dynamical viscosity, $\rho \nu$, is homogeneous, then the
viscous force can be simplified as
\begin{equation}
  \nabla \cdot \left( 2 \rho \nu {\sf S} \right)=
  \rho \nu \left( \bm{\nabla}^2 \bm {u} + \frac{1}{3} \bm{\nabla} \left( \DIVm \bm{u} \right) \right).
\end{equation}
The turbulent component of the velocity field can have its origin in
the cascade of energy from large scales towards small scales
thanks to the nonlinear interactions in the Navier--Stokes equation
\citep{K41}.  However in galaxies, deviating from the standard
picture
of forcing at largest scales driven, e.g., by large-scale shear,
energy injection occurs at multiple scales from
distributed stellar feedback in the form of ionising radiation,
stellar winds, and SNe; gravitational instability of the gas disc;
and, especially for young galaxies, gas accretion onto the
galaxy \citet{klessen2010}. Ionising radiation contributes an order of magnitude more
energy than the SNe \citep{abbott1982}, while SNe in turn contribute
an order of magnitude more energy than stellar winds
\citep{shull1995}.
The smallest contribution has been estimated to originate from the large-scale
shear, an order of magnitude less than the least powerful stellar source \citep{abbott1982}.
Stellar feedback causes mixing of the ISM at much
smaller scales than those associated with systemic large-scale flows
and ordered magnetic fields.

Conservation of mass implies
\begin{equation}
\frac{\pd \rho}{\pd t} + \bm{\nabla}\cdot(\rho\bm{u})=0,
  \end{equation}
while conservation of thermal energy then implies
\begin{equation}
\rho \frac{De}{Dt} = -p\DIVm{\bm u} + \bm{\nabla} \cdot \bm{q} + 2 \rho
\nu \vert {\sf S}\vert^2 + \eta \mu_0 \bm{J}^2 + \rho \Gamma - \rho^2\Lambda,
\end{equation}
where $\bm{q}$ is the conductive heat flux and the last four terms
describe the viscous and resistive heating and radiative heating and
cooling.  The radiative heating and cooling depend sensitively on the
temperature, ionisation state, chemical abundances, metallicity, and
dust properties (see \citealt{klessen2016} for a comprehensive review
of interstellar processes). The simplest approximation is to treat
heating as a constant rate due to far-ultraviolet ejection of
photoelectrons from dust and follow the cooling as a function of
temperature. More complete models include descriptions of radiation
transport and chemistry including both ionisation and molecule
formation to determine both the heating and the cooling.  The
temperature-dependent structure of the cooling function allows gas in
thermal equilibrium in an isobaric medium to exist at two temperatures
for typical interstellar pressures, which are described as thermal
phases \citep{field1969,wolfire1995,wolfire2003}.  Hot gas at low
density takes sufficiently long to cool that it acts as a third
quasi-stable phase \citep{mckee1977}.

The most common way of
closing the equations is to use the ideal gas law
\begin{equation}
  p = (\gamma - 1) \rho e,
\end{equation}
where $\gamma=c_{\rm P}/c_{\rm V}$, $c_{\rm P}$ is the heat capacity
at constant pressure, and $c_{\rm V}$ that at constant volume,
providing the relation between density and pressure required to
solve this set of equations.  However, note that $\gamma$ also varies with
the ionisation and chemical state of the gas, so in the general case
this equation is position dependent.

In addition to the stellar feedback, the ISM can undergo instabilities
of various kinds. Especially the magnetorotational instability (MRI),
gravitational instability, and the instabilities caused by cosmic rays are
considered to be important for galactic dynamo action.

The MRI \citep{BH91,piontek2004,piontek2005,piontek2007} can alter
both large- and small-scale dynamics. It has been semi-analytically
argued, however, that this instability would be suppressed by intense
SN activity up to a galactocentric radius of 15\kpc\ with Milky Way
parameters \citep{KKV10}, but it could still affect the dynamics
significantly in the outer regions \citep{SB99a} explaining the
anomalously high velocity dispersions there
\citep{tamburro2009,Beck15}.

At smaller scales, SN remnant shells undergo fragmentation, e.g., due
to the Richtmeyer-Meshkov instability (Rayleigh--Taylor instability due
to acceleration rather than gravity) or either of the Vishniac
instabilities \citep{vishniac1983,vishniac1994}.  These instabilities
have been argued to be important in 'breaking up' the shells of
superbubbles and enable them to overcome the flux-freezing constraint
and act favourably for a magneto-buoyantly driven dynamo \citep[see,
  e.g.,][]{Kulsrud2015}. The theory for this instability, dubbed as
the spike instability, has been developed in the early linear
stage. It seems unlikely that the assumptions applied in this stage,
such as treating the escape of magnetic flux as a diffusive process,
are valid in the non-linear stage, however.

Another way of triggering magneto-buoyancy in galactic disks is the
Parker instability powered by cosmic rays released in SN explosions
\citep{Parker92}.  The significant pressure exterted by the cosmic
rays enables the frozen-in magnetic field, even without fragmentation,
to buoyantly rise up to the galactic halo.  Meanwhile a portion of the
gas will sink along the magnetic field lines of the so-formed loops,
and finally form clumps at the connected footpoint of the flux
tubes. These magnetic loops can be sheared by differential rotation,
twist, and amplify the magnetic field \citep{HL97}.

Another instability induced by interstellar cosmic rays is the
streaming instability. According to the analytic theory
\citep{Bell2004}, positively charged cosmic rays, streaming along a
mean magnetic field, e.g., of LSD origin, induce constant currents
along supernova-driven shock jumps in the density of the cosmic
rays and background plasma. In the presence of a magnetic perturbation
perpendicular to the shock, the action of the Lorentz force causes the
background plasma to move transversely, while the cosmic ray plasma is
much less affected. If the magnetic perturbations are helical, these
motions can amplify the magnetic field by orders of magnitude
\citep[see, e.g.,][]{RS2009}. Even though the scale of the plasma is
much smaller than the Larmor radius of the background magnetic fields,
this instability has recently been argued to cause significant effect
on the scale of the small-scale magnetic field and interstellar
turbulence \citep{RS2009}.

Gas-rich galactic disks characteristic of high-redshift galaxies
  have strong gravitational instabilities of the type described by
  \citet{toomre1964} and \citet{goldreich1965}.  These drive
  turbulence from large scales independent of stellar feedback
  \citep[e.g.][]{krumholz2018} that drive prompt SSD action early in
  galactic history (see Sect.~\ref{res:cosmo} below).  Furthermore,
  \citet{klessen2010} showed that accretion flows, whether on to
    galaxies or individual star-forming cores, can drive substantial
    amounts of turbulence.  These flows have been shown by numerical
    models to be dynamo
    active in the case of first star formation
    \citep[e.g.,][]{sur2010} and during galaxy formation
    \citep[e.g.]{pfrommer2022}. For further review, see Sects.~\ref{res:sf} and~\ref{res:global} below.

\subsection{Basics and challenges of dynamo theory} \label{sect: dtheory}

In order to model dynamos, both the velocity and magnetic field can
naturally be decomposed into coherent and fluctuating parts, so that
we can write $\ut=\um + \uf$ and $\Bt=\Bm + \Bf$, and follow their
evolution separately. Ideally, the averaging should be over many
independent ensembles of the turbulent velocity
field, but in practice, spatial averages are used.  If the averaging
operation satisfies the Reynolds rules \citep[see, e.g.,][]{BS05},
then no strict requirement for clear separation between the average
and fluctuating scales exists, and the decomposition of the equations
into these two parts is generally valid.

As the largest scale galactic magnetic fields are predominantly
axisymmetric, azimuthal averaging obeying the Reynolds rules has
traditionally been used. This choice can be criticised on at least two
grounds. Firstly, many magnetic field configurations in galaxies are
not axisymmetric, so a Fourier filtering based decomposition would be
a more general choice for these. Secondly, observational diagnostics
are hardly ever derived basing on such averaging. Gaussian filtering
was successfully applied to decompose into the mean and fluctuating
fields by \cite{GSSFM13}, which would be a more consistent choice when
comparing with observations \citep[see, e.g., the discussion
  in][]{Hollins+22}.  The major difficulty with these alternative
averaging techniques is that neither of them obeys the Reynolds rules,
although this is no hindrance to deriving alternative, mathematically
sound formalisms for the mean quantities \citep[e.g.][]{Germano92}.
The bulk of analytic and mean-field galactic dynamo models continue to
use azimuthal averaging and, hence, we employ it in this review of the
theoretical concepts.

\subsubsection{Large-scale dynamo instability} \label{sect:LSD}

After taking the azimuthal average of the induction equation, we obtain an evolution equation for
the mean field
\begin{equation} \label{eq:MF}
\frac{\pd \Bm}{\pd t} = \bm{\nabla} \times \left(\um \times \Bm + \emf - \eta \bm{\nabla} \times \Bm \right),
\end{equation}
where $\emf = \overline{\uf+\Bf}$ is the turbulent electromotive force,
which depends on the correlations of the
velocity and magnetic fluctuations.
The evolution equation for the magnetic
fluctuations is obtained by subtracting the mean-field equation from the original induction
equation:
\begin{equation} \label{fluct}
\frac{\pd \Bf}{\pd t}=\bm{\nabla} \times \left(\um \times \Bf + \uf \times \Bm + \uf \times \Bf -\emf -\eta \bm{\nabla} \times \Bf \right).
\end{equation}
Two regimes are usually distinguished, where the Lorentz force
due to the mean magnetic field (1) is still weak, so the
magnetic field will not back react on the flow, called the {\em
  kinematic regime}; (2) is significant, so the magnetic field
influences the flow, called the {\em nonlinear regime}. Further
  distinction must be made in SSD-active systems, where also magnetic
  background turbulence will exist; under these circumstances, there
  exist magnetic fluctuations, ${\bm b}_0^{'} \neq 0$, unrelated to
  the presence of $\Bm$, and the turbulent transport coefficients can
  depend on these background flunctuations \citep[see, e.g.,][]{RB10}.

In the kinematic regime, further restricting to the case with
  ${\bm b}_0^{'}=0$, the presence of a mean magnetic field will
induce a fluctuating magnetic field through the term $\uf \times
\Bm$. Therefore, even if $\uf$ and $\Bf$ were uncorrelated without the
mean magnetic field, in its presence they will be correlated so that
$\uf \times \Bf$ is no longer zero, and in addition linearly
correlated with $\Bm$. If $\Bm$ varies over scales much larger than
those of the turbulent fluctuations, then one can expand $\emf$
in a Taylor series expansion of the form
\begin{equation}\label{emf}
\emfi=a_{ij} \Bmj + b_{ijk} \frac{\pd \Bmj}{\pd x_k}+ ...
  \end{equation}
and truncate after the first-order spatial derivatives of $\Bm$.
Here, $a_{ij}$ and $b_{ijk}$ are tensors that describe the influence
of turbulence on the evolution of the mean magnetic field, commonly
described as turbulent transport coefficients.  Particularly, the
symmetric part of the $\bm{a}$ tensor, usually denoted with
$\bm{\alpha}$, describes the collective inductive effect resulting
from turbulent motions, and the symmetric part of the rank two
  tensor acting upon ${\bm \nabla}\times\Bm$, usually denoted with
$\bm{\beta}$, describes the enhanced magnetic diffusivity resulting
from them.  In the kinematic regime, the transport coefficients may
depend on the key hydrodynamic and thermodynamic quantities, but not
on $\Bm$ itself.  With this simplification, only the evolution of the
magnetic fields at the largest scales can be captured. Simplifying
even further, as done in the oldest and most traditional first-order
smoothing approximation (FOSA) or second-order correlation approximation (SOCA)
\citep[see, e.g.][]{BS05} the term $\bm{\nabla} \times (\uf \times \Bf
- \overline{\uf \times \Bf})$ is neglected in Eq.~(\ref{fluct}) for
the fluctuating field. This is valid only when the magnetic
  Reynolds number, $\Rm\ll1$, or when the Strouhal number
\begin{equation} \mathrm{St}=\frac{\tau
  u}{l}\ll1,\end{equation} where $\tau$ is the correlation time of the turbulence,
and $u$ and $l$ are typical velocity and length scales,
respectively. Under
one of these assumptions, the equation for the fluctuating
field, Eq.~(\ref{fluct}), simplifies to the extent that it can be solved
analytically for a given $\uf$, after which $\emf$ can be calculated
with relative ease \citep{SKR66}.

Particularly, in the case of isotropic and homogeneous turbulence, the
turbulent transport coefficients then reduce to scalars,
\begin{eqnarray}
  \alpha \approx -\frac{1}{3} \tau \overline{\uf \cdot \omega'},\label{emf:FOSAalp}\\
  \beta \approx \frac{1}{3} \tau \overline{\uf^2},\label{emf:FOSAbeta}
\end{eqnarray}
where $\omega'$ is the fluctuating vorticity, and $\uf \cdot \omega'$ is the small-scale kinetic helicity
of the turbulent flow.
It can be seen that $\alpha$ relates to the kinetic helicity of turbulence,
and $\beta$ to the intensity of the turbulent motions.
A more generic form of the turbulent emf can be obtained by relaxing isotropy and homogeneity,
which certainly can
be appropriate
in galaxies, where rotation and large-scale shear are present:
\begin{equation}
\emf={\bm \alpha}\cdot\Bm+{\bm \gamma}\times\Bm
-{\bm \beta}\cdot(\nab\times\Bm)
-{\bm \delta}\times(\nab\times\Bm)
-{\bm \kappa} \cdot(\nab\Bm)^{(s)},
\label{emf:general}
\end{equation}
$\alpha$ and $\beta$ tensors now acquire off-diagonal elements. Completely new
effects arise, as discussed, e.g., by \cite{KR80}:
$\ggamma$ describes turbulent pumping of the mean magnetic field
\citep[e.g.,][]{OSBR02}; $\ddelta$ encompasses new dynamo effects arising
from current and rotation \citep[$\Omega \times \bm{\overline{J}}$ or R\"adler effect][]{Rae69} or
shear \citep[shear-current effect][]{RK04};
$(\nab\Bm)^{(s)}$ is the symmetric part of the derivative tensor of $\Bm$;
and $\kkappa$ is a rank-three tensor, which has also been shown to be important
in the dynamo process by \cite{WRVGTK21}, although it is neither fully diffusive nor
inductive.

In the nonlinear regime, still restricting to the case ${\bm
    b}_0^{'} = 0$, the turbulent transport coefficients may depend on
both $\Bm$ and on $\Bf$.  The outcome of taking into account these
nonlinearities is strongly dependent on the closure adopted for
Eq.~\eqref{fluct} \citep[see, e.g., ][]{SS22}.  Until the turn of the
millennium, most nonlinear galactic dynamo models adopted a heuristic
closure, where mainly the effect of $\Bm$ on the $\alpha$ effect is
postulated to be important, so that the inductive effect is suppressed
when the mean field is approaching equipartition with the turbulence:
\begin{equation}\label{aq:heuristic}
   \alpha=\frac{\alpha_0}{1 + \vert\Bm\vert/\vert B_{\rm eq}\vert}.
\end{equation}
Here, $\alpha_0$ is the kinematic value of the $\alpha$ effect and
$B_{\rm eq}\equiv \rho \overline{{\bm u'}^2}$ is the field strength at
equipartition with the turbulent kinetic energy.  Evidence from
numerical simulations in fully periodic domains in the beginning of
the 1990's \citep[e.g.][]{CV91,VC92} forced the dynamo community to
seriously consider other, more complicated and complete, closures,
some of which had already been formulated years ago,
\citep[e.g.][]{Pouquet76,VK83} and derive better nonlinear models
based on them.  Numerical simulations of forced turbulence, also
exhibiting exponential amplification of magnetic fluctuations through
the SSD (case of ${\bm b}_0^{'} \neq 0$), showed that the $\alpha$
effect is not quenched according to Eq.~(\ref{aq:heuristic}), but in a
catastrophic fashion, where also the magnetic Reynolds number appears
in the denominator. Practically, this was interpreted to imply that
galactic (or any astrophysical) $\alpha$-effect based dynamos should
not exist.

However, this quenching behavior was later understood not to result
from the SSD-generated fluctuations, but to be a consequence of
magnetic helicity conservation \cite[summarised e.g. by][]{BS05}:
under the mean-field approximation, a dynamo produces helicities of
equal magnitude but of opposite signs to satisfy helicity
conservation. If no helicity fluxes escape the dynamo active region,
then the small-scale magnetic helicity will quench the turbulent
electromotive force parallel to the mean field, leading to the
catastrophic quenching of the LSD. This is because the $\alpha$
effect acquires a magnetic modification, so that
\begin{equation}\label{eq:alpKM}
  \alpha \equiv \alpha_{\rm K} + \alpha_{\rm m}.
\end{equation}

One path into deriving the magnetic contribution to the
$\alpha$ effect is the so-called minimal $\tau$ approximation, where
the triple correlations appearing in the emf closure problem are
modelled as a relexation term \citep[for details, see,
  e.g.,][]{BS05}. The turbulent transport coefficients then read
for incompressible flows
\begin{eqnarray}
  \alpha = -\frac{\tau}{3} \left(\overline{\bm{\omega}' \cdot \uf} -
  \overline{\bm{j}' \cdot \Bf} / \overline{\rho} \right), \label{eq:taua} \\
  \beta = \frac{\tau}{3} \overline{\uf^2}, \label{eq:taub}
\end{eqnarray}
where ${\bm j}^{'} = {\bm \nabla} \times {\bm b}^{'}/\mu_{0}$ is
  the fluctuating current density, and $\overline{\rho}$ is the mean
  density. According to the magnetic helicity conservation law,
  the magnetic part has to satisfy \citep[see, e.g.,][]{BS05}
leading to an evolution equation
\begin{equation} \label{eq:alpdyn}
  \frac{\pd \alpha_{\rm m}}{\pd t}= -\eta_{\rm t} k_0^2 \left( \frac{2 \emf \cdot \Bm + \nabla \cdot \overline{{\bm{F}}}}{B_{\rm eq}^2} +\frac{2 \alpha_{\rm m}}{\widetilde{\Rm}}\right).
\end{equation}
Here $\eta_{\rm t} = \eta+\beta$ is the sum of microscopic
resistivity, $\eta$, and the enhanced turbulent resistivity, $\beta$,
$\overline{\bm{F}}$ is the large-scale helicity flux of the
  fluctuating field, $k_0$ is the wavenumber corresponding to
  the integral scale of turbulence, and $\widetilde{\Rm} \equiv
  \eta_{\rm t}/\eta$.  The most advanced mean-field (MF) models for
galaxies solve this equation together with the induction
equation. With some plausible estimates from observations or theory
for the coefficients of the models, galactic MF dynamos remain useful
tools \citep[see, e.g.,][]{SS22}, but verifying their applicability
rigorously remains an open question (see the discussion in
Sect.~\ref{MF:results}).

\begin{figure}[ht]
\centering
  \includegraphics[width=0.9\textwidth]{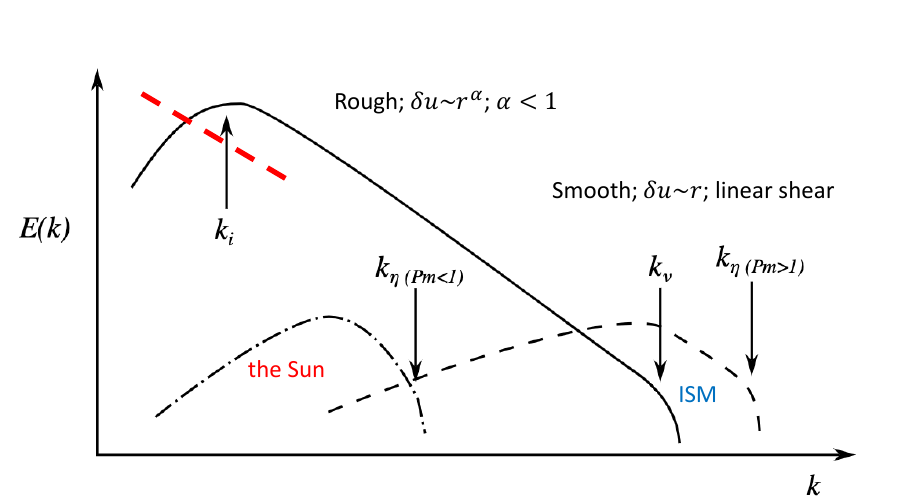}
  \caption{Schematic illustration of different types of SSD
    spectra. The solid line represents the kinetic energy spectrum,
    with a cut-off scale $k_{\nu}$, $\nu$ being the kinematic
    viscosity. The dashed lines represent magnetic energy spectra in
    the kinematic stage, with cut-off scales $k_{\eta}$, $\eta$ being
    the magnetic resistivity. Two different scenarios, differentiated
    by the magnetic Prandtl number, $\Pm=\nu/\eta$, are shown,
    corresponding for low (such as the Sun) and high (such as the
    ISM). The red long-dashed line illustrates the contribution of a
    LSD at the low wavenumber range. Image adapted from
    \cite{RBBRKL23}.   \label{fig:spectrum}}
\end{figure}

\subsubsection{Small-scale dynamo instability} \label{sect:SSD}

The ISM of galaxies is, in general, a high magnetic Prandtl number
fluid, meaning that the ratio of magnetic to fluid Reynolds numbers
\begin{equation}
  \Pm\equiv\frac{\nu}{\eta}=\frac{\Rm}{\Re},
\end{equation}
is large;
\cite{BS05} estimate values of $\Pm \sim \mathcal{O}(10^{11})$.  The
resistivity, $\eta$, is then much smaller than the fluid viscosity,
$\nu$, meaning that the dissipation scale of the fluid motions far
exceeds that of the magnetic field. This implies that the magnetic
fields can dissipate into much smaller-scale structures than the fluid
motions themselves, and hence a typical magnetic field structure sees
the velocity field as a smooth, larger-scale shearing flow acting on
it, (compare spectra plotted with solid and dashed lines plotted in
Fig.~\ref{fig:spectrum}).

In such a setting, a rapid amplification of magnetic fluctuations by
the smooth fluid motions can occur, independent of the flow possessing
helicity or not. This dynamo instability is called the SSD or
fluctuation dynamo. The scenario was first discussed by
  \citet{bachelor1950}, and the first rigorous treatment of the
non-helical case was presented in \citet{K68}. The
amplification occurs at the timescale of the turbulent eddy turnover
time, which is far shorter than the timescale required for the
amplification of the large-scale field by an LSD.  In the
kinematic growth phase, the structures generated have a thickness
around the resistive scale, but are curved up to the scale of the
turbulent eddies.  The only requirement for this instability to
set in is that $\Rm$ must exceed a critical value, analytically
derived to be 30--60, a condition that is fulfilled from the integral
to the dissipation scale in galaxies \citep[see e.g.][]{Kandu98}.

Presumably, the setting in which the SSD starts to operate in galaxies
is one where the LSD has not yet produced a significant large-scale
field, but battery mechanisms and plasma instabilities can produce
seed fields at small scales in the regime of dissipation scales rather
than within the inertial range (see Sects.~\ref{subsec:seeds}
and \ref{res:plasma}).  Therefore, the SSD excitation likely starts in
a diffusion-dominated setting, in contrast to the other possible, but
unlikely, scenario of SSD starting with larger-scale tangled fields,
in the diffusion-free regime \citep{Rinconrev19}.  In the kinematic
stage, when the magnetic fluctuations are still weak, the peak in the
magnetic power spectrum is at the resistive scale, but all scales grow
with the same growth rate.  At scales larger than the resistive scale,
the magnetic power spectrum is predicted to follow a positive power
law of $k^{3/2}$ \citep{K68}, called the Kazantsev spectrum.  At
smaller scales, the spectrum falls off steeply, following the
Macdonald function \citep[for details, see e.g.][]{BS05,Rinconrev19}.
For a schematic representation of the magnetic spectrum, see
  Fig.~\ref{fig:spectrum} dashed line.

How the SSD nonlinearly saturates remains an open problem. By
extending the Kazantsev model, it has been proposed that the SSD grows
magnetic fluctuations at the resistive scale up to and exceeding
equipartition with the kinetic energy of the turbulence, but that the
generated field would be concentrated into resistive-scale ropes,
hence not being volume filling, and therefore being energetically
insignificant \citep[e.g.][]{Kandu98}.  Whether reconnection plays a
role in the nonlinear regime of SSD is a relatively fresh research
question arising from such numerical research that is discussed in
this review.

  In galaxies that have already excited LSD, another source of
  magnetic fluctuations comes through tangling of the coherent
  component by the action of the turbulent eddies. Currently, the
  fraction of fluctuations caused by the SSD versus tangling is not
  known observationally, and calls for numerical approaches.
  Ideally, models where both SSD and LSD instabilities and their
    non-linear interactions and saturation regimes are realistically
  captured are required to study the relation of the regular and
  fluctuating components. Such modelling requires large
    scale separation ratios, and thereby high resolution, and the
    models need to be integrated over timescales of gigayears.
  Currently, numerical models reach only modest magnetic Prandtl
    number regimes, while extending to the extreme regimes in galaxies
    is yet impossible and likely to remain that way in the near
    future. Even when working in parameter regimes far-removed
    from the ISM, numerical simulations with modest Prandtl numbers
  have provided some answers, and here we review the current
  state-of-the-art.

\subsection{Seed fields}
\label{subsec:seeds}

Both SSD and LSD need initial magnetic fields to grow, as their
evolution equations are homogeneous in $\Bm$ and $\Bf$,
  respectively. There have been a number of mechanisms proposed to
generate such seeds. These have included seeds of cosmological origin
\citep[see, e.g., the review by][]{Durrer13}; battery mechanisms
\citep[see, e.g.][]{Kandu94}; and stars picking up the battery fields
during star formation, amplifying them through a stellar dynamo, and
then ejecting the fields as seeds in SN explosions \citep[][]{Sasha88,
  MartiAlvarez21}.  However, recent work has demonstrated conclusively
that kinetic plasma instabilities can generate magnetic fields from
field-free initial conditions that in the presence of turbulence grow
rapidly with an SSD, as we describe below in
Sect.~\ref{res:plasma}.
These novel findings alleviate the
  previous concerns of the origin of strong enough seed fields to be
  grown by LSD alone in galaxies.  Therefore, we leave further
detailed discussion of the origin of seed fields out of the current
review.

\section{Numerical approaches} \label{sect:numerics}

\subsection{Mean-field modelling} \label{MF:models}

The simplest numerical models use the approximations discussed in
Sect~\ref{sect:LSD} to study the evolution of the large-scale
component of the galactic magnetic field.  Furthermore, as the
rotation laws of galaxies are well-constrained observationally, the
models most often work in the limit where the large-scale velocity
field is steady instead of being evolved according to the mean-field
Navier--Stokes equation.  As such dynamo models do not solve for the
full dynamical problem, but only for the evolution of the mean
magnetic field (Eq.~\ref{eq:MF}), they need ad hoc parameterisations
of the turbulent transport coefficients. A comprehensive review of the
MF approach and the results with given differential rotation
  profiles has been recently presented in \cite{SS22}. We
  summarise their theoretical discussion in the following, while for
  details and the results in the limit of given differential rotation
  laws we refer to their Chapters 11 and 12.

The dominant turbulent effects in galaxies are assumed to be the
$\alpha$ effect and the enhanced turbulent diffusion, while turbulent
pumping and the more exotic dynamo effects discussed in
Sect.~\ref{sect:LSD}, are usually regarded to be sub-dominant.
Furthermore, the scalar expressions for isotropic and homogeneous
turbulence, Eqs.~\eqref{emf:FOSAalp} and \eqref{emf:FOSAbeta}, are
often assumed to be valid.  As discussed earlier, it is
observationally well-established that strong magnetic fluctuations
dominate over the mean component in galaxies. The effect of the
fluctuations, most likely implying strong pre-existing magnetic
background turbulence ($b_{0}^{'} \neq$ 0), are currently not taken
into account through the turbulent transport coefficients as such, but
rather through their non-linear saturation formulae,
Eqs.~(\ref{eq:alpKM})--(\ref{eq:alpdyn}).

Observations allow for estimates of the magnitudes of $\alpha$ and
$\beta$, and also that of the amplifying effect of differential
rotation, making galactic dynamos better constrained than many other
astrophysical dynamos, such as the solar one.  Nevertheless,
confirmations from numerical models, not working under the MF model
assumptions, are required to test the validity of the various
assumptions used, and to fill in details that the observations can not
reveal due to the limited accuracy and projection effects.  Such
attempts, tools used, and the results obtained are discussed in
Sect.~\ref{TFM:results}.

Cylindrical coordinates ($r,\phi,z$), with the $z$-axis aligned with
the rotation vector, and $(r,\phi)$ being coordinates in the
horizontal plane, are the most often used coordinates to simulate the
MF equation.  Under the assumption of axisymmetry, $\pd/\pd \phi$
vanishes. If we can further assume that the disc is very thin, that
is, $\epsilon=h_0/R \ll 1$, where $h_0$ is the scale height and $R$ is
the galactocentric radius, one may also neglect all the terms
with radial derivatives, so the evolution equation for $B_z$ decouples
from that of the horizontal components.  One is then left with
the problem of solving the evolution equations for the radial and
azimuthal components.  In this approximation, the MF equations contain
only $z$-derivatives, but the turbulent transport coefficients and the
mean field itself can depend parametrically on the radius, and
are thus local.  The non-dimensional MF equations contain two control
parameters for the LSD, called dynamo numbers:
\begin{equation}
R_{\Omega}=\frac{S_0 h_0^2}{\beta}, {\ \ } R_{\alpha}=\frac{\alpha_0 h_0}{\beta},
  \end{equation}
where '0' subscripts now refer to characteristic values determined,
e.g., from observations, and $S$ describes the shear, for flat galactic
rotation curves $S=-V_0/R$ being a good approximation, with $V_0$ being the
roughly constant rotational velocity.
Their product $D=R_{\alpha} R_{\Omega}$
characterises the overall dynamo efficiency.  If we adopt the
so-called $\alpha \Omega$ assumption, where the azimuthal magnetic
field is solely generated by differential rotation, neglecting a term
arising from the $\alpha$ effect, that can be estimated to be much
smaller, then the parameter $D$ is the only non-dimensional control
parameter describing the full problem. Using qualitative arguments,
that is, not solving any dynamo equations yet, the growth rate
$\lambda$ is expected to scale as
\begin{equation}
\lambda^{-1} \approx \frac{h_0^2}{\beta}| D |^{-1/2}.
  \end{equation}
For numerical solutions, vacuum boundary conditions are often
used, the justification coming from the estimate that the magnetic
diffusivity beyond the disc is
one to two orders of magnitude higher than in it, and the
large-scale magnetic field is largely confined to the thin disc.

The assumption of axisymmetry is a strong restriction, as many
galaxies have been observed to host non-axisymmetric magnetic field
configurations (Sect.~\ref{sect:obs}).  The approach outlined above
can, however, be relatively easily applied to a non-axisymmetric
system as well. Now the parametric dependence of the mean magnetic
field not only includes $R$, but also the
azimuthal angle $\phi$. The turbulent transport coefficients can be
either axi- or non-axisymmetric.  There is one major difference to be
accounted for: differential rotation acts differently on axi- versus
non-axisymmetric solutions, efficiently destroying the latter, while
the $\alpha$ effect does not make such a strong distinction.  Hence,
the $\alpha \Omega$ approximation is not applicable in this regime,
and both dynamo numbers must be kept in the problem. Again, for very
thin discs, an asymptotic solution can be constructed, where
radial derivatives vanish, and the evolution equation for the
$z$-component decouples.  Qualitative considerations show that the
main characteristics of the dynamo solutions, such as the growth rate,
should remain invariant with respect to the axisymmetric
solutions. Only numerical studies reveal differences, \citep[see, e.g., ][]{SS22}.

The fundamental process through which galactic dynamos are expected to
saturate is magnetic helicity conservation, leading to the dynamical
$\alpha$ effect, Eq.~\eqref{eq:alpdyn}. Galactic discs, although often
treated in MF models as thin discs, interact strongly with their
environments, with stellar feedback driving galactic fountains and
winds and material accreting from the surrounding circumgalactic
medium \citep[e.g.][]{tumlinson2017}.  Under such conditions, magnetic
field is expected to be transported outwards when tied to the hot,
ionised gas in fountains and winds, with some of it falling back with
cooler and denser structures in the fountains and during
accretion. Also, a turbulent pumping effect may be important here in
transporting mean field inwards or outwards. All these processes will
result in re-distribution of magnetic helicity at different scales,
and helicity fluxes between different regions. Parameterising these
effects in terms of helicity fluxes from the existing numerical models
is a tremendous challenge. What is currently assumed in the MF models
is described in Sect.~\ref{MF:results}, and what is known from
numerical simulations is discussed in
Sect.~\ref{HelicityFluxes:results}.

\subsection{Direct numerical simulation versus large eddy schemes}

Nonlinear numerical simulation of dynamo action requires that
resistivity be added, either explicitly or by discretisation error, to
the equations of ideal MHD.  The discretisation error present in any
numerical method does provide a nonlinear resistivity that can enable
or even suppress dynamo activity. Theoretically, an LSD can be excited
whenever $\Rm>1$, while an SSD is known to require higher threshold
values of order of tens to hundreds (see Sects.~\ref{sect:LSD} and
\ref{sect:SSD}). Therefore, an LSD should be easier to model
numerically than an SSD, but an LSD requires many more pre-requisites
from turbulence, such as the standard expectation of small-scale flows
being helical, to be satisfied than an SSD.

Including explicit diffusivities in numerical models at the extremely
small values characteristic of the ISM, while retaining their correct
ratio as reflected by the extremely large $\Pm$ in the ISM, is
currently an unreachable challenge. Often models including explicit
diffusivity are called direct numerical simulations (DNS), but this
denotion is not strictly valid due to the orders of magnitude elevated
diffusivities required to exceed numerical diffusivity. A more proper
term of DNS-like models has recently been proposed
\citep{RBBRKL23}. To be useful, such implementations must be run at
high enough resolution that the numerical resistivity does not
dominate over the physical values chosen.  A parameter study to find
the resolution where the physical resistivity starts to determine the
solution must be done to demonstrate that numerical resistivity is not
dominant.

Using explicit diffusivities has the disadvantage that it provides
constant diffusion everywhere, while the necessity for using diffusion
is the largest at the grid scale to ensure numerical stability and
  allow magnetic field lines to reconnect, while no such need exists
for the resolved scales. Hence, implicit large-eddy simulations (ILES)
are often used, where no explicit physical diffusion terms
are included, but numerical diffusion remains active.
In astrophysical modeling, the most usual sources of numerical
  diffusion are provided by
  truncation error in Riemann solvers \citep[e.g.]{balsara2004} or hyperdiffusivities introduced
  for numerical stability
  \citep{brandenburg2002,brandenburg2020}. \citet{grete2023} recently
  showed that such ILES schemes produce results that compare surprisingly well to the results of
DNS-like models with well-resolved dissipation scales.
The disadvantage of these sorts of ILES schemes is that
  control parameters such as $\Re$, $\Rm$, and $\Pm$ become
  ill-defined. To counter-act this shortcoming, measurements have been performed of numerical
resistivity \citep{rembiasz2017,mckee2020,bian2021} and of numerical viscosity
\citep{schranner2015,obergaulinger2020,mckee2020,bian2021} by comparing results
of models without physical diffusivity to analytical or numerical
solutions including physical diffusivity.
Despite these efforts, the ILES approach
remains unsatisfactory, as the diffusivity has a
steep, unphysical scale dependence, with all diffusivity focused at
scales close to the grid.

Explicit large-eddy simulations (ELES), where the solved equations are
filtered at an intermediate scale, and the smaller-scale effects are
described through a (ideally) physically motivated sub-grid-scale
(SGS) model, have been developed for idealized MHD turbulence
\citep{Smagorinsky63,grete2017}.  Although SGS models to model
processes as SNe and star formation are popular, numerical SGS models
attempting to describe the dynamo processes in the ISM remain
scarce. One recent example is the model by \cite{liu2022} that
includes an unresolved turbulent dynamo. The most complicated case for
the ELES schemes is presented by an LSD, in which small-scale,
unresolved, effects on the large-scale dynamics, usually referred to
as back-scatter in the LES framework, should be captured by the SGS
model. Steps in the direction of a true ELES MHD model for supersonic
turbulence have been taken by \citet{grete2015} and \citet{grete2017},
but such an SGS model has not yet been incorporated into galactic
dynamo simulations.

\subsection{Spectral methods}

The first published numerical simulation of a turbulent dynamo was performed
with a spectral method \citep{meneguzzi1981}.  Spectral methods have
the highest accuracy derivative per grid point in smooth flows
\citep[e.g.][]{maron2008}, but struggle to simulate the
super-Alfv\/enic flows characteristic of the interstellar medium
because of Gibbs ringing at shocks, so they have not featured
prominently in studies of galactic dynamos.

\subsection{Grid methods}

\subsubsection{Uniform grid methods}
Discretisation of the MHD equations onto a uniform Cartesian grid
provides the most straightforward numerical approach.  For the dynamo
problem, a prime advantage of uniform grid methods available in codes
such as the Pencil Code \citep{BD02,brandenburg2020}, Athena \citep{stone2008},
Athena++ \citep{stone2020}, {\sc ZEUS-MP} \citep{hayes2006}, {\sc
  NIRVANA-III} \citep{ziegler2004,ziegler2011}, or Piernik
\citep{hanasz2010a,hanasz2010b} is the uniform numerical dissipation
properties across the computational domain. This advantage is
unfortunately lost in the non-Cartesian geometries available in these
codes, where different grid zones have different sizes or aspect
ratios.  \citet{kritsuk2011} compares single-grid algorithms for
isothermal, magnetised, supersonic turbulence showing that algorithmic numerical
diffusivity and resolution can be traded against each other to get
equally good solutions.

\subsubsection{Hierarchical grid methods}
Adaptive mesh refinement (AMR) methods focus resolution in subgrids
into regions of interest, typically where strong gradients occur.
\cite{kritsuk2006} demonstrated in isothermal, supersonic,
hydrodynamic turbulence without an explicit viscosity using Enzo
\citep{collins2011} that if the driving and dissipation scales are
well separated, and the base grid has sufficient resolution to resolve
the larger scales of the turbulent cascade, then AMR of dissipation
regions can reduce the cost of the simulation without changing the
statistical properties of the turbulence.  \cite{li2012} similarly
demonstrated using the {\sc ORION2} AMR MHD code that the statistics
of magnetised turbulence can equally well be modeled with AMR methods,
though with an emphasis on the properties of the highest density,
highest resolution regions at the cost of the low density regions.

AMR methods have been applied to SSDs in dwarf galaxies by
\cite{rieder2016} and \cite{rieder2017}, and placed in a
cosmological context in \cite{rieder2017a}.
\cite{martin-alvarez2022} argue that to study SSDs in cosmological
simulations of galaxies, refinement down to a uniform grid
encompassing the galactic disc yields better results than a refinement
strategy strictly tied to density gradients that results in lower
resolution in low-density warm regions.

\subsection{Particle methods} \label{method:particle}
There are multiple methods tied to particles rather than a structured
grid that have been used for dynamo simulations, including smoothed
particle hydrodynamics (SPH), meshless finite mesh (MFM) and meshless finite
volume (MFV) methods, and Riemann solvers on a Voronoi mesh.

\cite{dolag2009} implements MHD in the GADGET SPH code
\citep{springel2005} using a \cite{powell1999} eight-wave divergence
cleaning algorithm, which diffuses magnetic divergence across the grid
to constrain it at a low level.  This was used by subsequent
publications for study of galactic field evolution in isolation or
cosmological context
\citep[e.g.][]{beck2012,beck2013,steinwandel2019,steinwandel2020}.  An
Euler potential approach \cite{price2007} is implemented by
\cite{kotarba2009} into the VINE SPH code \citep{wetzstein2009},
providing excellent conservation of field divergence but constraining
magnetic helicity unphysically.  \cite{wissing2020} implements MHD
within the geometric density averaged force algorithm used by the
Gasoline2 SPH code \citep{wadsley2017}.  This was used by
\cite{wissing2022} to model dynamo activity in an isolated disc
galaxy. However, these methods remain problematic because of the lack
of zero-order convergence \citep{morris1996,dilts1999,read2010} and
the large kernel sizes needed to run modern versions stably.

\cite{pakmor2013} describes the implementation of MHD in the Voronoi
mesh code AREPO \citep{springel2010} using an approximate Riemann solver to derive fluxes
across the mesh edges. They used a \cite{powell1999} scheme to maintain field
divergence small. This was used by
\cite{pakmor2014} and \cite{pakmor2017} to simulate galactic dynamos
in cosmological context.  \cite{mocz2016} showed how to use
constrained transport \citep{evans1988} with AREPO, to maintain zero divergence to
machine precision even on the unstructured mesh.

An alternative approach to modelling fluid flow with particles is
offered by weighted meshless schemes \citep{gaburov2011}, as
implemented in the MFV and
MFM methods used in GIZMO \citep{hopkins2015a}. These methods use
values at a particle and its neighbors to fit a polynomial to the solution and its
derivatives, rather than averaging over the values and then taking
derivatives using differences of the averages as SPH does. The
difference between the two GIZMO methods lies in the decision of
whether the Riemann problem at zone faces should be solved in the
frame of motion of the fluid at the face (MFM) or at the mean cell
velocity (MFV).  MHD is
implemented with a Godunov HLLD solver and a \cite{dedner2002} magnetic
divergence damping scheme.

\subsection{Hybrid and kinetic methods} \label{method:hybrid}

Dynamos in collisionless plasmas, such as those found in the
intracluster medium surrounding galaxies or in the hot early universe, can not be properly modeled with the
MHD approximation because of the presence of fast kinetic instabilities.
\cite{rincon2016} used a hybrid model that treats protons as kinetic particles and
electrons as an isothermal fluid to study this problem.
\citet{pusztai2020} used a fully kinetic model implemented with a
discrete Galerkin solution of the kinetic Boltzmann equation limited
to subviscous scales for computational tractability.
\cite{sironi2023} and \cite{zhou2023} describe fluctuation dynamos modeled with a
particle-in-cell treatment of an electron-positron pair plasma.   These
methods are vital for understanding the origin of seed fields, but
are less relevant for most flows within the bulk ISM of galaxies.
Extending them to include electron-ion mass ratios and greater dynamic
range in scale will be essential for making contact with the physical
parameter ranges occurring during galaxy formation.

\section{Results from numerical approaches}

Simulations of dynamo action in galaxies have demonstrated its
importance at all scales.  Many expectations from analytical theory
are supported, and the detailed field structure and evolution can be
traced in this highly nonlinear problem. However, limitations of
numerical resolution and the different numerical approximations used
can result in qualitatively different outcomes in various situations,
necessitating careful evaluation of reported results.

\subsection{Plasma scale} \label{res:plasma}

The generation of magnetic fields by cosmic accretion shocks from
unmagnetised plasma was hypothesised by \citet{gruzinov2001} based on
consideration of the properties of gamma ray burst
shocks. \cite{schlickeiser2003} and \cite{lazar2009} made analytic arguments for the
\cite{weibel1959} instability in shocks being the physical mechanism
that could generate these fields when the shocks drive streams of
electrons and ions moving in different directions. These ideas were
supported by initial models of homogeneous Weibel instability performed with
particle-in-cell codes by \cite{medvedev2004} and \cite{sakai2004}.

\citet{rincon2016} demonstrated that magnetic field formation from an
unmagnetised turbulent plasma and subsequent dynamo growth can occur
using a hybrid model that includes an electron fluid and proton
particles computed using a full six-dimensional Vlasov equation.  This
model suppresses both the \cite{biermann1950} battery and
\cite{weibel1959} instability because of its assumption of isothermal electrons.
However, \cite{pusztai2020} used their fully kinetic model to
demonstrate that Landau damping can limit the growth of kinetic
dynamos in some regimes, an effect that cannot be seen in the hybrid
approximation.

\begin{figure}[ht]
\centering
  \includegraphics[width=0.9\textwidth]{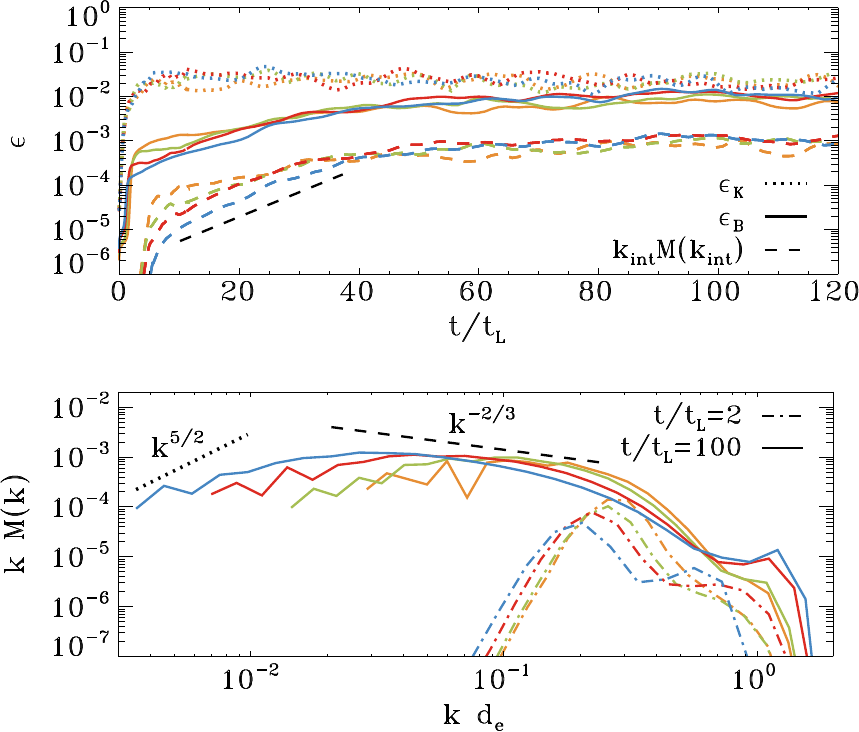}
  \caption{Kinetic models of an initially unmagnetised turbulent
    flow. \textbf{Top} Magnetic energy $\epsilon_B$ {\em (solid lines)}, kinetic energy
    $\epsilon_K$ {\em (dotted lines)}, and magnetic energy at the
    turbulent driving scale $k_{\rm int}$. Time is given in units of the crossing time
    $t_L$ at
    $k_{\rm int}$ for models with domain sizes $L$ in terms of the
    electron skin depth $d_e$ of $L/d_e$ ranging from 250 {\em (orange)} to 2000 {\em (blue)}.
    The dashed black line shows exponential growth of the rms field
    $B_{\rm rms} \propto \exp(0.4v_{\rm rms} t / L)$. \textbf{Bottom} Magnetic
    power spectra immediately after magnetisation by Weibel
    instability {\em (dot-dashed lines)} and in the quasisteady state after dynamo saturation
    at equipartition {\em (solid lines)}, with the Kazantsev ($M(k)
    \propto k^{3/2}$ and
    Kolmogorov $(M(k) \propto k^{-5/3}$ spectra shown in {\em black}. Image reproduced with permission from \cite{sironi2023}, copyright APS.}
  \label{fig:sironi-etal-23-f3}
\end{figure}

By including both positively and negatively charged particles in a fully
kinetic simulation of a hot pair plasma that excludes the \citet{biermann1950}
battery effect, \citet{sironi2023} and \citet{zhou2023} show that magnetic
fields generated by the \citet{weibel1959} instability from an unmagnetised
plasma can grow by SSD in driven turbulence, as shown in
Fig.~\ref{fig:sironi-etal-23-f3}.

The separation between driving and dissipation scales in these kinetic
models remains far smaller than in galactic or intergalactic gas, and
models still need to be done with more appropriate ratios of electron
to ion masses.  Nevertheless, the mechanisms described seem robust
enough that we should conclude that such an SSD will occur already
behind the accretion shocks that occur during the assembly of gas into
galaxies.  As a result, magnetic fields of at least a few percent of
equipartition with turbulence likely provide the initial condition for
{\em any} galactic dynamo.  Simulations starting with extremely small
seed fields may unphysically delay the onset of magnetohydrodynamic
effects in galactic evolution.

\subsection{Star formation scale} \label{res:sf}

During the formation of the first stars, dynamo action can already occur in
the turbulent accretion flows assembling the gas into haloes. Kinetic
simulations of this process were discussed in
the previous section. \cite{schleicher2010} presented analytic
estimates of the resulting field strengths in halos, showing that
near equipartition fields could already be expected even prior to the formation of the
first stars.

After the accreted gas cools sufficiently to produce an
accretion disc surrounding a protostar, magnetorotational instability
in this disc can drive an SSD \citep{BNST95}. A semi-analytic model of
the SSD in such a disc by \cite{schober2012} established that fast
growth would be expected.  This treatment was extended to quickly accreting young
galaxies by \cite{schober2013}.

\begin{figure}[ht]
  \centering
  \includegraphics[width=0.7\textwidth]{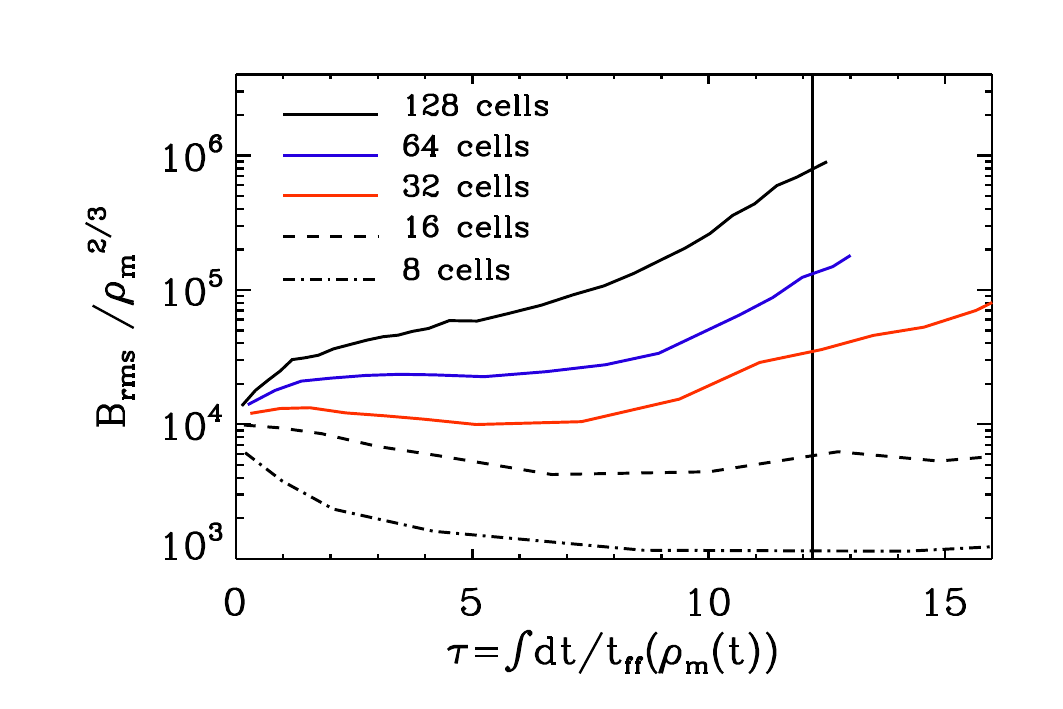}
  \caption{The number of cells with which the Jeans length is resolved
    in a primordial star formation region determines whether growth of
    the magnetic field from an SSD can be resolved.  Amplification by
    spherical gravitational collapse produces field growth
    proportional to $\rho^{2/3}$, so further field growth is evidence
    of the action of the SSD.
    Image reproduced with permission from \cite{sur2010}, copyright vy AAS.}
  \label{fig:sur-etal-2010-f4}
\end{figure}

After the action of an SSD during accretion treated in the previous
section, dynamos in protostellar
accretion discs are another source of macroscopic fields on short time
scales during galaxy formation.   Ideal AMR MHD models with Flash \citep{fryxell2000} by
\citet{sur2010} and \citet{federrath2011} demonstrated that an SSD can only be resolved
in an accretion disc if the \cite{truelove1997} criterion for
resolving self-gravitating discs is satisfied with at least 32
cells, as shown in Fig.~\ref{fig:sur-etal-2010-f4}.

 In this case, exponential growth can occur with timescales
shorter than the orbital time of the disc.  \cite{mckee2020} and
\cite{stacy2022} show growth of fields in primordial discs to values
characteristic of modern star forming regions in simulations starting
with cosmological structure formation and zooming in to disc scale.
By characterising the numerical diffusivity, they found that the
growth rates were consistent with the non-ideal MHD SSD theory of
\cite{xu2016}.

Whether modern protostellar accretion discs maintain their magnetic
fields through the dynamo action of the MRI depends on how ionised
they are.  Dead zones where ambipolar or Ohmic diffusion dominate have
fields in balance with magnetocentrifugally-driven winds, as suggested
by local shearing-box models with Athena \citep{bai2013}, and then
confirmed in a global disc model on a spherical-polar mesh using {\sc
  NIRVANA-III} \citep{gressel2015}.  Active zones sufficiently ionised
for the gas to fully couple to the magnetic field indeed have
MRI-driven SSD action, as shown \citep{BNST95} using uniform-grid, shearing-box,
open-boundary simulations with the Stagger code \citep{nordlund1990,brandenburg1996}, and in
stratified, shearing-box simulations \citep{stone1996} with the {\sc
  ZEUS} code \citep{stone1992,stone1992a}.

\subsection{Kiloparsec scale} \label{res:kpc}

The structure of the magnetic field within the ISM on the kiloparsec
scale can be separated into turbulent scales and the large-scale or
MF. The separation scale is given by the effective 75--200 pc forcing
scale of the turbulence driven by SN blast waves
\citep{joung2006,avillez2007} and stellar winds from massive stars
\citep{vanBuren85}.  While the blastwave of a single SN evolves
over a range of scales from its stellar radius to tens of parsecs, the
effective forcing scale for the ISM turbulence is controlled by the
collective action of multiple SNe combining to blow superbubbles
\citep{Korpi:1999a, joung2006, avillez2007, gent2013,
  HSSFG17}. Small-scale drivers from low-mass stellar winds,
protostellar jets, planetary nebulae, and cloud dynamics also exist,
but these are negligible compared to SNe and, to a lesser extent, the
highest mass stars \citep{mac-low2004}.  The resulting turbulence can
support a turbulent dynamo (or SSD), which amplifies the magnetic
field exponentially.

The MF is organised by processes such as the differential rotation of
galactic discs, spiral arms, and other metagalactic scale processes
\citep{BS05,SS22}.  ISM turbulence, can, however, also be driven by
large scale processes such as gravitational instability
\citep{rafikov2001}, Kelvin-Helmholtz and other gas-dynamical
instabilities, disc-halo circulation, and Parker instability.

\subsubsection{SN driven SSD}\label{kpc:ssd}
Theoretical approaches to modelling the magnetic field in the ISM commonly
presume approximate equipartion between the energy density contained within
each of the thermal, kinetic, cosmic ray and magnetic field energy.

However, numerical models of SSD action in the ISM of galaxies
suggest that it produces and
maintains magnetic fields of only a few percent of equipartition with the
kinetic energy.  This low saturation of the kinematic dynamo is evident for
isothermal simulations at high Mach number \citep[e.g.,][]{Haugen:2004M,
FCSBKS11, FSBS14}.  Models of turbulence simulated at the value of
$\Pm>1$ relevant in the ISM show that the level of saturation increases with Pm, but still appears to reach an
asymptotic limit well below equipartition \citep[e.g.,][]{SBK02, Schober12,
SBSW20}.

With isothermal turbulence driven by pure solenoidal forcing
\citep{BFKMS22} it is possible to reach around one third equipartition
for $\Pm=4$ at $\Re=500$.  \cite{seta2020} and \citet{BFKMS22} show
that the SSD grows independently of the properties of the seed field
in individual galaxies, while \cite{garaldi2021} comes to the same
conclusion using self-consistent cosmological simulations.  SSD in a
two-phase medium grows more slowly than in isothermal turbulence of
comparable Mach number and Rm \citep{seta2022}.  It does appear that
the SSD alone is insufficient to account for the observed strength of
the turbulent magnetic field in galaxies.

\begin{figure}[ht]
\centering
  \includegraphics[width=0.8\textwidth]{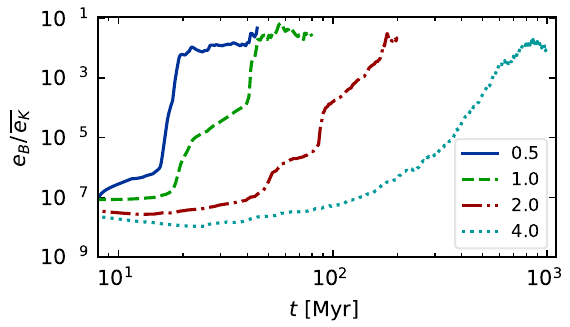}
  \caption{Magnetic energy scaled by time-averaged kinetic energy
    in a periodic box driven by SN explosions and radiative
      heating and cooling
    to produce a turbulent multiphase medium including explicit
    Laplacian resistivity and viscosity. Numerical resolution in
    parsecs is given in the legend. The timescale of exponential
    growth drops dramatically as resolution improves and Re$_{\mathrm{M}}$
    increases. Image reproduced with permission from \citet{GMKS21}, copyright by AAS.}
  \label{fig:gent-etal-21}
\end{figure}

With more realistic forcing in a multiphase ISM by SNe, at rates more relevant
to spiral galaxies, but without any drivers for LSD,
\cite{balsara2004,BalKim05} and \citet{GMKS21} also find that the SSD saturates at a few
percent of equipartition strength, but much faster than in
isothermal models or the cooler two phase
medium, as shown in Fig.~\ref{fig:gent-etal-21}. Because of the stochastic
nature of the driving, this dynamo acts intermittently \citep{GMKS21}.  Field
growth is fastest in the hot phase where high sound speed facilitates high
vorticity \citep{gent2023}, but the strength at which the SSD saturates is lowest
in the hot gas and increases in cooler phases.  In particular, the turbulent
flows of diffuse, inhomogeneous, hot gas in SN remnants and
suberbubbles host the most rapid SSD \citep{gent2023} and \citep[][Fig.~1,
  third row]{KMG22}.

In elliptical galaxies lacking differential rotation there is no driver for LSD, but
SSD remains active, even though SN rates are much lower than in disc
galaxies. The theoretical model of \cite{MS96} anticipates a turbulent
magnetic field similar in magnitude to those of spiral galaxies, while
\cite{SRFH21} conclude from simulated observations of numerical
turbulence that turbulent magnetic field must be an order of magnitude
weaker.

Due to the random distribution of SNe and the multiphase structure of the ISM,
the turbulent properties are highly anisotropic and intermittent. A thorough
investigation of dependence of the turbulent magnetic properties on temperature
phase and SN rate, and on galactic radial and vertical anisotropy remains
outstanding. This would form the basis for modelling simulated observations of
the magnetic structure of the ISM as well as constructing effective
subgrid-scale models of MHD turbulence with which to explore large-scale
magnetic effects.

\subsubsection{Large-scale fields without LSD}\label{kpc:imp}

To achieve the field strength observed in spiral galaxies, and also to
comprehensively understand the anisotopy, structure and strength of the
turbulent magnetic field, requires the presence of an LSD.

Some investigations have explored the evolution of a large scale field
in a stratified ISM
\citep[e.g.,][]{deAvillez:2005,Hill:2012a,walch2015,girichidis2016},
but with an imposed initial uniform MF, which cannot be sustained in
the absence of an MF dynamo mechanism. The turbulent field in these
models arises primarily through tangling of the initial field.
Comparisons between \cite{balsara2004} and \cite{GMKS21} indicate that
at the resolution applied in such models and using these codes, the
SSD would take some 100 Myr to initiate and up to a gigayear to
saturate, so most of these models are likely still in the
  unsaturated growth stage of the SSD.

\subsubsection{SN driven LSD}\label{kpc:lsd} 

To solve directly the evolution of the MF dynamo with
turbulence driven by SN explosions \cite{Korpi:1999b} and \citet{Korpithesis99} model a
stratified disc subject to rotation using a shearing box.  Neither the SSD nor the
LSD could be sustained because the model did not have the required
resolution to reach magnetic
Reynolds numbers sufficient to excite the SSD. With open vertical boundaries
and a relatively low vertical extent the gas in these models is rapidly
exhausted and overheated within a few hundred megayears.  This is too soon to amplify
the MF dynamo, which it is reasonable to expect be primarily hosted in
the continuous warm phase of the ISM, rather than the isolated cold filaments
or hot bubbles.

With the Nirvana code and double the grid resolution
\cite{gressel2008} and \cite{Gressel08br} model an area 800 pc on a
side extending $\pm 2.1$~kpc either side of the galactic
disc. Exploring angular velocities ranging from $1\OSN$ to $8\OSN$,
where the Sn subscript denotes the Solar neighborhood value, with
and without shear and with SN explosion rates between $0.25\ZSN$ and
$1\ZSN$, they find the LSD to be effective for
$\Omega\gtrsim2\OSN$. In the absence of Coriolis forces the shear
produces no LSD. The efficiency of the LSD increases slightly as SN
rate reduces. The LSD does not saturate in these results, but it is
clear that the timescale would exceed 1.5 Gyr for $\Omega \leq 4\OSN$.

\cite{KO15B} using a similar setup, but with the inclusion of a shearing
radial boundary at
$1\OSN$, evolve their
models beyond 850 Myr. Where the initial field is strong enough, the turbulent
tangling and any residual SSD are sufficient to sustain an equilibrium total
magnetic energy throughout, but for the model with a weak initial field the
final state remains far below equipartition. This setup has the essential
ingredients to model the LSD, but given the absence of any clear exponential
growth it is very difficult to differentiate SSD from tangling and to exclude the
effects of the imposed field on the results.

With the Pencil Code, \cite{gent2013} almost double the resolution again to
model a domain of horizontal area $(1.024 \mathrm{ kpc})^2$ and extending
$1.088$~kpc either side of the disc.  Applying differential rotation via the
shear-periodic boundary, they confirm an LSD for
$\Omega\geq1\OSN$. For a model with $\Omega=2\OSN$ they obtain an MF
dynamo with a growth rate of approximately 5\,Gyr$^{-1}$, about twice
that for the
growth rate of the turbulent field. This growth, however, measures only the
period 800\,Myr to 1.05\,Gyr, well after the SSD has saturated, but while the
LSD remains an order of magnitude less than equipartition.

From comparison of the evolving mean and fluctuation magnetic energy, it is
evident that the SSD is absent for \cite{gressel2008} while it is very active
for \cite{gent2013}. It is not clear why this would impede the progress of the
LSD at low $\Omega$ for the former, while not in the latter. The LSD should be
quite insensitive to the Rm in these ranges. It poses the question as to the
extent the SSD may hinder or assist the action of the LSD. \cite{gent2013} do
not obtain magnetic energy spectra with an obvious separation of scales between
the mean and fluctuating magnetic fields, which instead show a single parabola.
They do, however, show that a meaningful separation of scales can be defined
with the fluctuating field occupying scales typically smaller than 250 pc and
the large scale field having scales in excess of 700 pc.

\begin{figure}[ht]
  \resizebox{0.5\textwidth}{!}{
    \includegraphics*[trim=-2.5cm 0cm 0cm 0cm,clip=true]{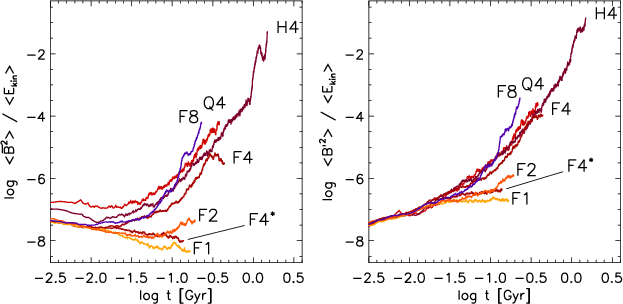}
  }\hspace{1.0cm}
  \resizebox{0.305\textwidth}{!}{
    \includegraphics*{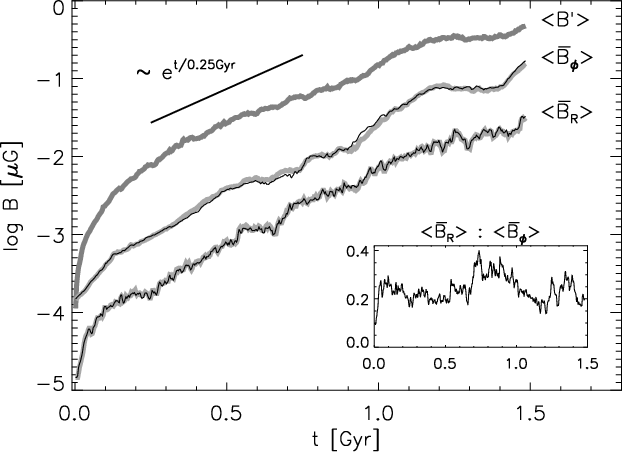}
  }\\
  \resizebox{0.45\textwidth}{!}{
    \includegraphics*{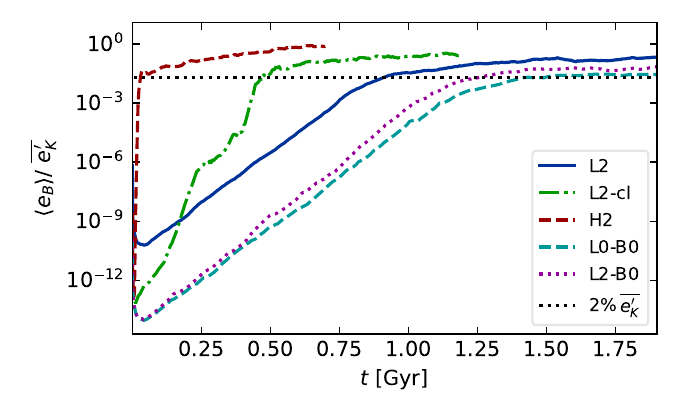}
  }\hspace{1.1cm}
  \resizebox{0.45\textwidth}{!}{
    \includegraphics*{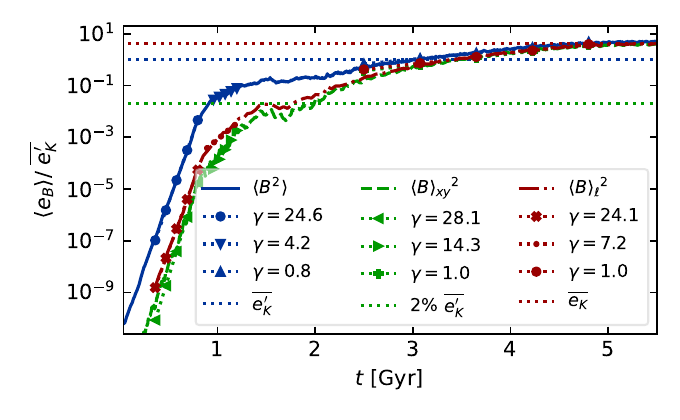}
  }
  \caption{For all models of \citet{gressel2008} {\em (a)} mean and {\em (b)} turbulent
magnetic energy evolution.  Labels Q, H and F denote SN rates of 0.25, 0.5 and
$1 \ZSN$, respectively, while the numbers give the angular
  velocity scaled by $\OSN$.  {\em (c)} Evolution of the
horizontally averaged and the turbulent magnetic field strength for model H4
and {\em (d)} evolution of total magnetic energy for all
models from \cite{Gent23SSDLSD}, where L denotes resolution of $\delta
x=4$\,pc and H $\delta x=1$\,pc. The suffix `cl' denotes use of SN clustering and
`B0' the ongoing
exclusion of an MF to isolate SSD activity. In both sets the numbers in the model
labels denote $\Omega$ scaled by $\OSN$. {\em (e)} Evolution of the total
magnetic energy in model L2 and the MF energy
calculated from horizontal averages and convolution with a Gaussian
kernel. Figure adapted from \citet{gressel2008} and
\citet{Gent23SSDLSD}.
\label{fig:gressel-v-gent}
}
\end{figure}

Figure~\ref{fig:gressel-v-gent} compares the models of the galactic LSD by
\cite{gressel2008} and \cite{Gent23SSDLSD}.
Figure~\ref{fig:gressel-v-gent}{\em (a)} shows that for SN rate $\sigma=\ZSN$ the growth
rate of the MF is proportional to $\Omega$ for $\Omega\geq2\OSN$.
\cite{gressel2008} do not find an MF dynamo for $\Omega=\OSN$. However,
\cite{Gent2012} finds an MF dynamo is present at $\Omega=\OSN$.  Model
F1 from \citet{gressel2008} extends only to about 150 Myr, though, which may have been too short to dissipate
transients from the initial azimuthal seed field.  The linear dependence of
the growth rate on $\Omega$ is consistent with the theoretical expectation of
\cite{BS05} for a supercritical MF dynamo, which suggests the
MF dynamo will be well supported in fast and even moderately rotating
galaxies where the rotation curve is flat.  In the range
$0.25\ZSN\leq\sigma\leq\ZSN$ the LSD is anticorrelated with $\sigma$.

The interaction of the SSD and LSD is explored by
\cite{Gent23SSDLSD}.  With the growth and saturation properties of the SSD
better defined by \cite{GMKS21,gent2023}, the effect of an active LSD on the
SSD is isolated by systematically removing any horizontal MF from the
components of the magnetic field in a stratified ISM with and without
differential rotation. In the case with additional energy injected by the large scale
shear from the differential rotation, the SSD grows slightly faster and saturates proportionately
higher than in
the non-rotating system (compare the growth rates of L0-B0 and
  L2-B0 in Fig.~\ref{fig:gressel-v-gent}{\em (d)}). When the LSD is not excluded in the differentially
rotating model (L2 in Fig.~\ref{fig:gressel-v-gent}{\em (d)},
  which does start with a higher seed field),
the SSD has a slightly lower growth rate during the kinematic phase. The LSD
extracts energy from the SSD equivalent to the additional energy injected by
the large scale shear. The turbulent field, however, does not stop growing after
the SSD would saturate in the absence of the LSD, though growth slows as the LSD
transitions into the dominant mode of the dynamo. The effect of the SSD on the
LSD requires further study.

From Fig.~\ref{fig:gressel-v-gent}{\em (c)} it can be seen that,
even
though \cite{gressel2008} does not show SSD activity, the turbulent field rapidly increases to around ten times
the strength of the seed field. This is supported by the analysis of
\cite{Gent23SSDLSD} during the period where the SSD is present, but saturated. The saturation
state of the SSD is confirmed by models L0-B0 and L2-B0 shown in
Fig.~\ref{fig:gressel-v-gent}{\em (d)}, where the MF is continuously
removed.  When the SSD saturates for model L2 (see
Fig.~\ref{fig:gressel-v-gent}{\em (e)}), the turbulent field continues
to grow due to the presence of the LSD even though it is over ten times
stronger than the growing MF which it is continuing to generate.
Tangling by SN turbulence therefore can produce
a turbulent field an order of
magnitude stronger than the MF.

\begin{figure}[ht]
\centering
  \resizebox{0.95\textwidth}{!}{
    \includegraphics*[trim=-0.65cm 0cm 0cm 0cm,clip=true]{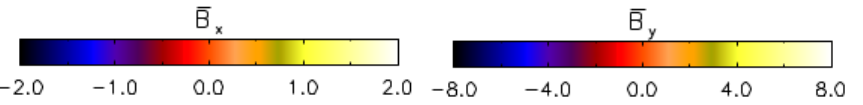}
  }
  \resizebox{0.95\textwidth}{!}{
    \includegraphics*[trim=-0.1cm 0cm 0cm 0cm,clip=true]{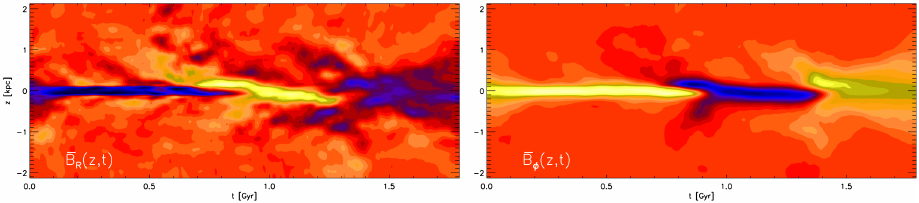}
  }
  \resizebox{0.49\textwidth}{!}{
    \includegraphics*[trim=0.45cm 0.45cm 0.65cm 0.25cm,clip=true]{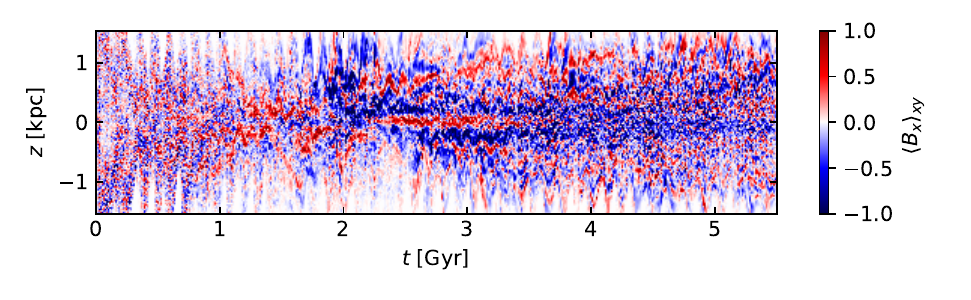}
  }
  \resizebox{0.49\textwidth}{!}{
    \includegraphics*[trim=0.45cm 0.45cm 0.65cm 0.25cm,clip=true]{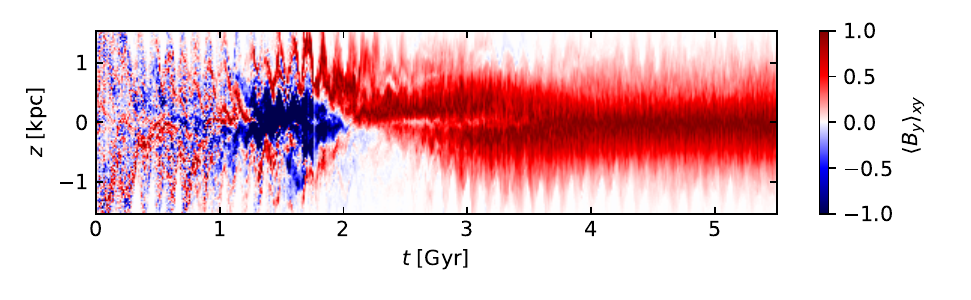}
  }
 \begin{picture}(0,0)
    \put(-320,107){{\sf{(a)}}}
    \put(-163,107){{\sf{(b)}}}
    \put(-320, 35){{\sf{(c)}}}
    \put(-151, 35){{\sf{(d)}}}
 \end{picture}
  \caption{Evolution of the horizontally averaged mean magnetic field
components, normalised by the maximum field strength at each time.
From model H4 of \citet{gressel2008} the field components {\em (a)}
$B_r$ and {\em (b)} $B_\phi$ are shown.
From model L2 of \cite{Gent23SSDLSD} the field components  {\em (c)}
$B_x$ and {\em (d)} $B_y$
are shown.  Both models
  are seeded by fields with maximum rms strength at the midplane $<
  1$~nG, and grow to strengths of several microGauss.
    \label{fig:gressel-v2-gent}}
\end{figure}
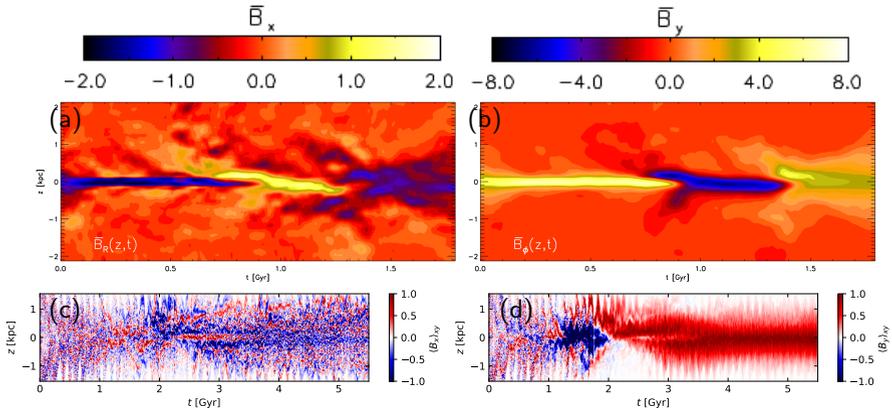

To understand how we might expect the dynamo generated mean field to appear
near the galactic plane, we compare in Fig.~\ref{fig:gressel-v2-gent}
time-latitude diagrams of horizontal averages for model H4 from
\cite{gressel2008} and L2 from \cite{Gent23SSDLSD}.  Model H4, in
panels {\em (a)}
and {\em (b)}, exhibits a quite regular field, subject to reversals with a period
below 1 Gyr, and has quadrupolar structure.  By contrast, model L2, in panels (c) and (d),
initially has a rapidly fluctuating structure both in time and latitude due to
the presence of a vigorous SSD.  Well after the saturation of the LSD, the
field becomes more regular. Early on there is more asymmetry with numerous
reversals between different latitudes.  As the field becomes more quadrupolar
there is then a reversal around 2~Gyr, after which the field saturates to a
steady azimuthally dominant field.  The comparsion between the models is most
relevant up to about 2 Gyr, when both models are still subequipartition. It is
noteworthy that the scale height of the magnetic field expands in the late
stages when the strength of the midplane magnetic field becomes dynamically
significant with respect to the kinetic energy.
Whether the presence of reversals in the field is impacted by the rate
of rotation, which differs by a factor of two here,  the scale height of the
disc, the SN rate, or the horizontal size of the domain requires
further investigation.

\subsubsection{Magnetic buoyancy instability}
Another source of dynamo instabilities explored at these scales is the magnetic
buoyancy instability (MBI) or when attenuated by cosmic rays, the Parker
instability.  A suitably stratified disc, with sufficiently strong parallel
magnetic fields, has been
initialized to examine the linear stage of the instability in an
isothermal galaxy by \cite{RSSBF16}. In such cases the mean magnetic
field is imposed, presumably acquired by prior galactic dynamo,
and the exponential growth of the velocity purturbations driven by the
MBI drives a secondary dynamo.  The resulting magnetic field has an
expanded scale height and can attain superequipartition strength
\citep{RSSBF16}. The growth rates during the linear stage
of the instability obtained from the analytic linearised
solutions of \cite{GS93} are of order 20--30 Gyr$^{-1}$.
Numerical solutions yield somewhat lower growth rates.
When the numerical ratio of magnetic to gas pressure is high
the growth rate of the linear instability is enhanced.

Including solid body rotation is anticipated analytically to dampen
MBI growth \citep{Ryu_2003} but \cite{KRJ01} find it makes little
difference to the growth of the linear instability, though it can
enhance the formation of turbulent structure in the magneitc field.
The nonlinear state of MBI has been examined in the context of disc
galaxies without \citep{Devika23a} and with \citep{Devika_Parker23}
rotation.  MBI is characterised by an $\alpha$-effect with opposite
sign of helicity to that typically driving the $\alpha\Omega$-dynamo
\citep{Devika_Parker23,QSTGB23}.  When rotation is included
\citep{QSTGB23} MBI induces periodic reversals in the large-scale
field \citep{Devika_Parker23} and the cosmic ray energy density is
higher.  Where the large-scale magnetic field is generated via a
dynamo, the MBI has little effect on the stratified structure of the
gas and cosmic rays, while still expanding the vertical scale height
of the magnetic field, whereas when the field is imposed, it expands
the scale height of the field and cosmic rays so far as to diminish
their pressure support, resulting in a thinner gas disc during the
nonlinear stage \citep{QSTGB23}.  In \cite{QSTGB23} an
$\alpha^2$-dynamo is applied to a nonrotating disc, so the effect of
differential rotation and the resulting $\alpha\Omega$-dynamo remains
to be explored, as does the interaction between SN-driven turbulence
and the MBI.

\subsubsection{Magnetorotational instability}

Investigating MRI-driven turbulence in a two-phase stratified ISM,
\cite{piontek2005} find that, even without SNe, MRI is sufficient to excite
strong turbulence in the diffuse ISM away from the midplane or in the outer
galaxy where the ISN in the disc becomes more diffuse.
MRI inhibits self-gravitating and thermal instabilites, likely retarding star
formation in the outer galaxy \citep{piontek2007}.  While turbulence can
increase the proportion of the ISM in thermally unstable temperatures, the
bimodal character of the ISM persists, consistent with the findings of
\cite{GSSFM13} for the trimodal character in the presence of SN-driven
turbulence. The strength of the magnetic field becomes somewhat independent of
gas number density \citep{piontek2005,piontek2007}.
However, decomposing the magnetic field between mean and turbulent contribution
\cite{EGSFB16} shows the turbulent magnetic field to correlate more with
the hot phase and the MF with the warm and cold phases.

\subsection{Galactic scale} \label{res:global}

\cite{schober2013} estimate the growth rate for an accretion-driven
dynamo in a spherical protogalaxy to be
\begin{equation} \label{eq:Gamma}
  \Gamma \simeq k v_k \mathrm{ Re}^{1/2}
\end{equation}
for wavenumber $k$, velocity at that scale $v_k$, and Reynolds number
Re, where the exponent is appropriate for Kolmogorov turbulence.  They
use this to analytically estimate that an accretion-driven small-scale
dynamo can amplify a weak seed field to near-equipartition with the
turbulent kinetic energy in tens to a few hundred megayears.  These
are almost certainly overestimates of the time required for these
fields to grow, since equipartition level small scale fields already
grow in the first protostellar accretion discs, as described in
Sect.~\ref{res:sf}.

Models of field growth in modern galaxies that start with small seed
fields can therefore only be interpreted as physics experiments
exploring the behavior of dynamos.  However, they can not be
intepreted fruitfully to understand the astrophysical evolution of
galaxies in the presence of magnetic fields, because of the extended
period of weak field present in these models that does not occur in
any scenario including cosmological growth of galaxies.

\subsubsection{Mean-field (MF) models} \label{MF:results}

There exist proof-of-concept non-linear MF models employing the full
dynamic quenching formula with various types of helicity fluxes. For
example, the studies of \cite{Shukurov06} and \cite{Sur2007a}
investigate galactic fountain-driven fluxes and the latter study also
investigates helicity fluxes resulting from anisotropic turbulence due
to the presence of shear, a scenario proposed by \cite{VC2001}. Also
diffusive helicity fluxes have been included by
\cite{Chamandy2014}. All these studies were able to demonstrate that
including helicity fluxes results in LSD saturating at the
equipartition value with turbulence being excited and
maintained. Also, evidence for the heuristic nonlinearity
(Eq.~\eqref{aq:heuristic}) producing results consistent with the more
complete quenching models was found \citep[e.g.][]{Chamandy2014}.
This is reassuring, as both old and new MF models still often use the
heuristic nonlinearity \citep[e.g.][]{liu2022}.

\subsubsection{MHD models} \label{subsub:gal-mhd}

\begin{figure}[ht]
\centering
  \includegraphics[width=0.9\textwidth]{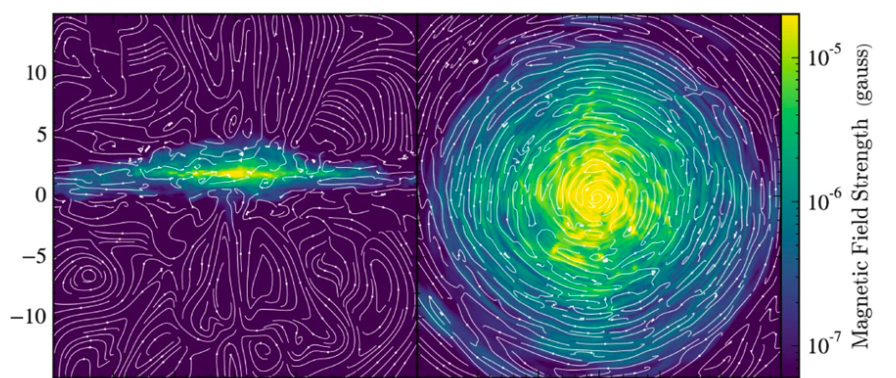}
  \caption{Magnetic field strength projected edge-on and face-on in an
    isolated disc galaxy. Directed stremalines of magnetic field are
    shown in white.   The length scale is given in kiloparsecs on
    the left axis.  Image reproduced with permission from \cite{butsky2017}, copyright by AAS.}
  \label{fig:butsky-etal-17-f2}
\end{figure}

The interrelation between the SSD and the LSD is a key question at the
galactic scale.  The physical timescale for the SSD is far shorter
than for the LSD.  However, insufficient numerical resolution can
dramatically slow the action of the SSD \citep{gent2023}, causing the
two timescales to apparently overlap in numerical simulations.
Nevertheless, numerical simulations do produce toroidal fields that
generally reproduce the observed topology, as shown, for example, in
Fig.~\ref{fig:butsky-etal-17-f2}.

\paragraph{Small-scale dynamo}

\begin{figure}[ht]
\centering
\includegraphics[width=\textwidth]{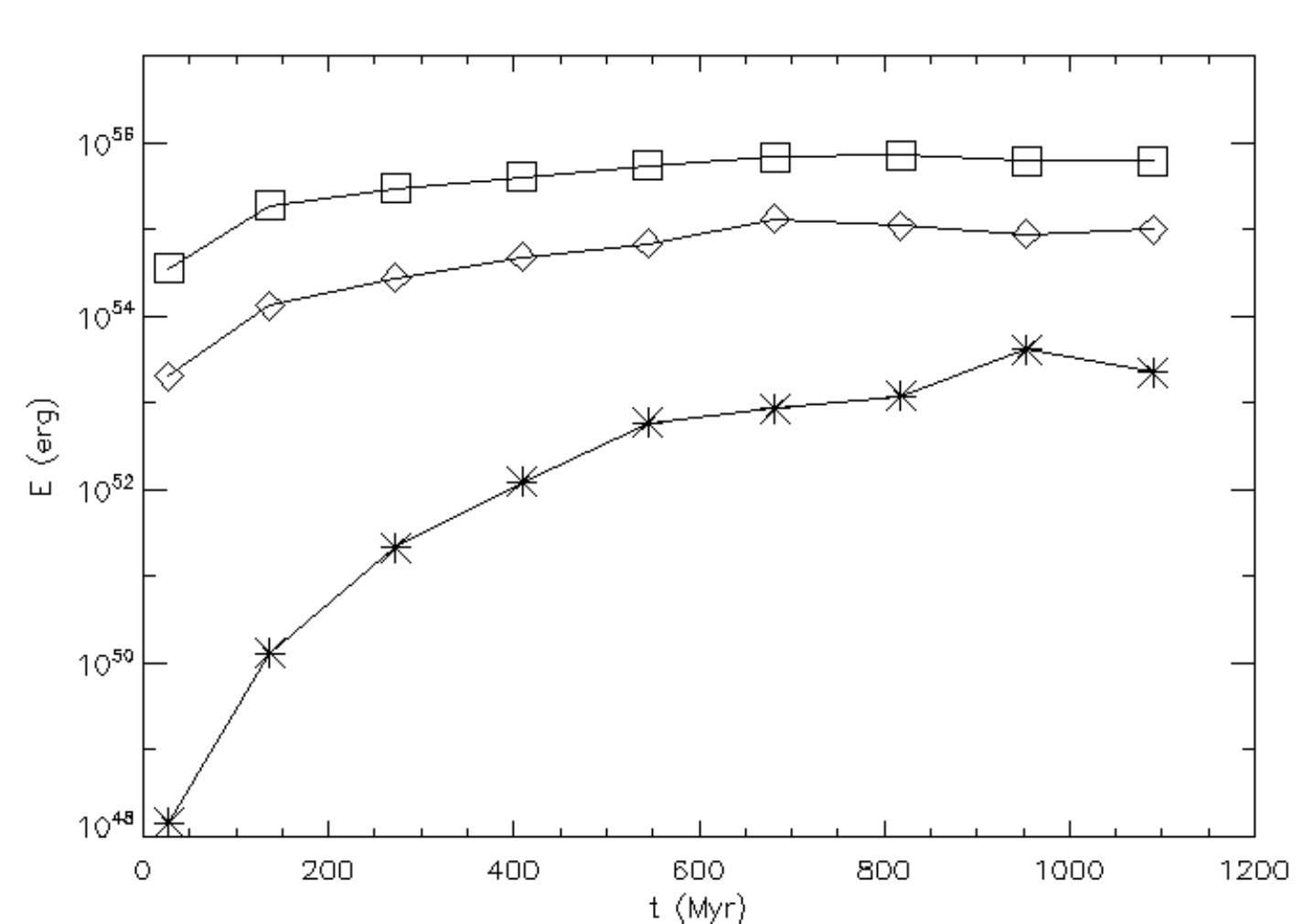}
  \caption{Magnetic energy (asterisks) in disc for a dwarf galaxy simulation at
    redshift two without stellar feedback at 26 pc resolution compared
    to kinetic energy (squares) and thermal energy (diamonds).
    including SN feedback at the finest resolutions given in
    the legend. Exponential growth with the given time scales is shown by
    the black lines. Image reproduced with permission from \cite{wang2009}, copyright by AAS.}
  \label{fig:wang-abel09}
\end{figure}

The presence of SSD in global galaxy models has been demonstrated
using two primary diagnostics: the exponential growth of magnetic
energy and the agreement of the power spectrum of the field with the
$k^{3/2}$ prediction of \cite{K68}. The first global MHD models of
field development in a disc galaxy were performed by \cite{wang2009}
and \cite{kotarba2009}, neglecting turbulence driven by stellar
feedback.  These works already demonstrated that the turbulence
generated by gravity and differential rotation in galaxies is
sufficient to produce exponential field growth to a few percent of
equipartition, as shown in Fig.~\ref{fig:wang-abel09}, although at a
rate limited by the low numerical resolution and turbulence strength.
\cite{kotarba2009} further demonstrated that constraint of magnetic
helicity by evolution of the field using Euler potentials
\citep{price2007} suppresses dynamo activity.

The first diagnostic, exponential growth of magnetic energy to a few
percent of kinetic energy in global galaxy simulations has been
confirmed by multiple groups in the last decade. Models without
explicitly modeled feedback-driven turbulence include
\cite{pakmor2013, rieder2016, pakmor2017,rieder2017,steinwandel2020}
and \cite{pfrommer2022}. In most of these models, feedback is assumed
to contribute to a turbulent pressure variable but the only flows
contributing to an SSD are produced by gravitational dynamics of the
galactic gas or large-scale galactic winds. A number of studies
without explicit feedback did not explicitly report the exponential
growth of magnetic energy, but rather that of related but not
identical quantities such as the average magnetic field strength,
including \cite{pakmor2014, steinwandel2019} or the median magnetic
energy density \cite{pakmor2017}. Models that do include stellar
feedback, driving turbulence at smaller scales and resulting in faster
magnetic energy growth, include \cite{su2017,butsky2017,ntormousi2020}
and \cite{martin-alvarez2022}.

\begin{figure}[ht]
  \centering
  \includegraphics[width=0.8\textwidth]{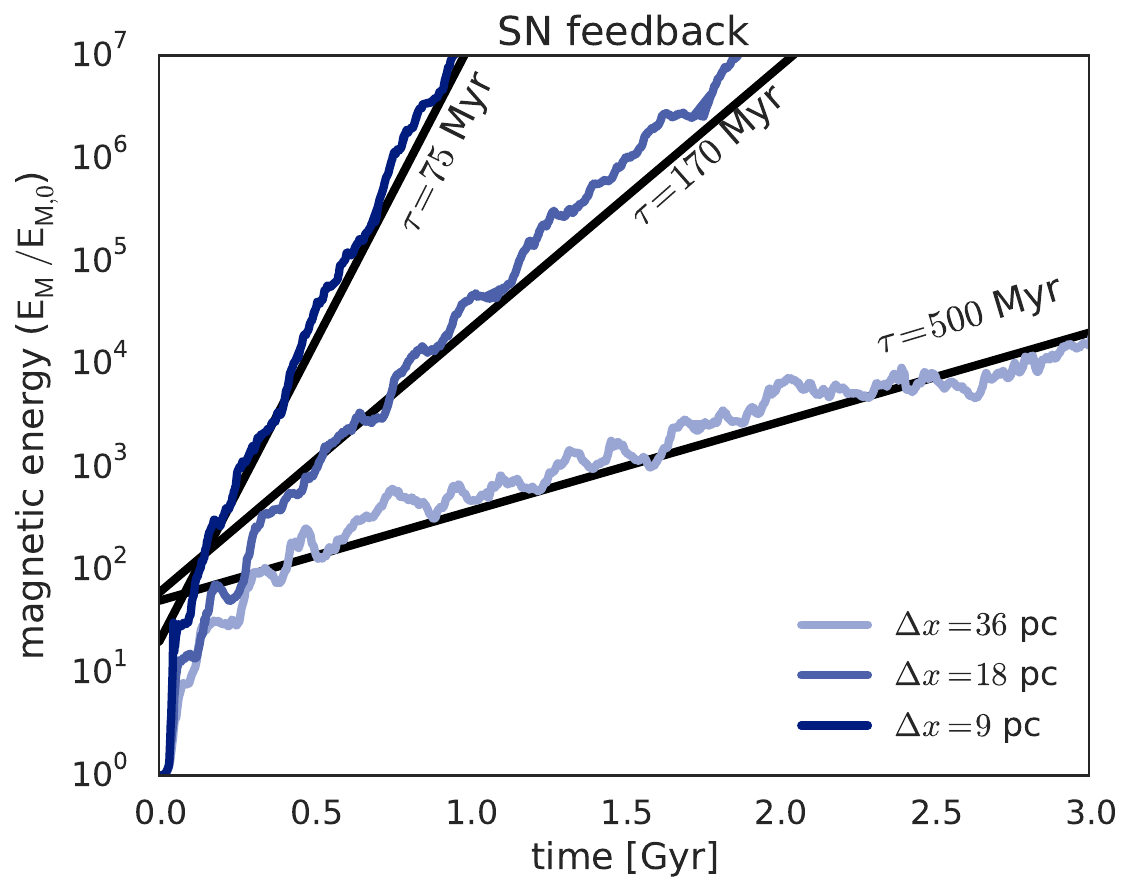}
  \caption{Magnetic energy in domain for a dwarf galaxy simulation
    including SN feedback at the finest resolutions given in
    the legend. Exponential growth with the given time scales is shown by
    the black lines. Image reproduced with permission from \cite{rieder2016}, copyright by the author(s).}
  \label{fig:rieder16-f9R}
\end{figure}

\cite{rieder2016} clearly demonstrated
the importance of numerical resolution in full galaxy models, as shown
in Fig.~\ref{fig:rieder16-f9R}.  Note that these are still low
resolutions and thus slow growth rates compared to those reached in
kiloparsec-scale models (see Sect.~\ref{kpc:ssd}), where growth time
scales drop to under a megayear (Fig.~\ref{fig:gent-etal-21} at
0.5~pc resolution. \citet{martin-alvarez2022} emphasised the
importance of uniformly resolving galactic discs rather than allowing
resolution to follow density to correctly capture the growth of
magnetic fields in lower-density regions where they grow faster.

\begin{figure}[htp]
  \centering
    \includegraphics[width=0.8\textwidth]{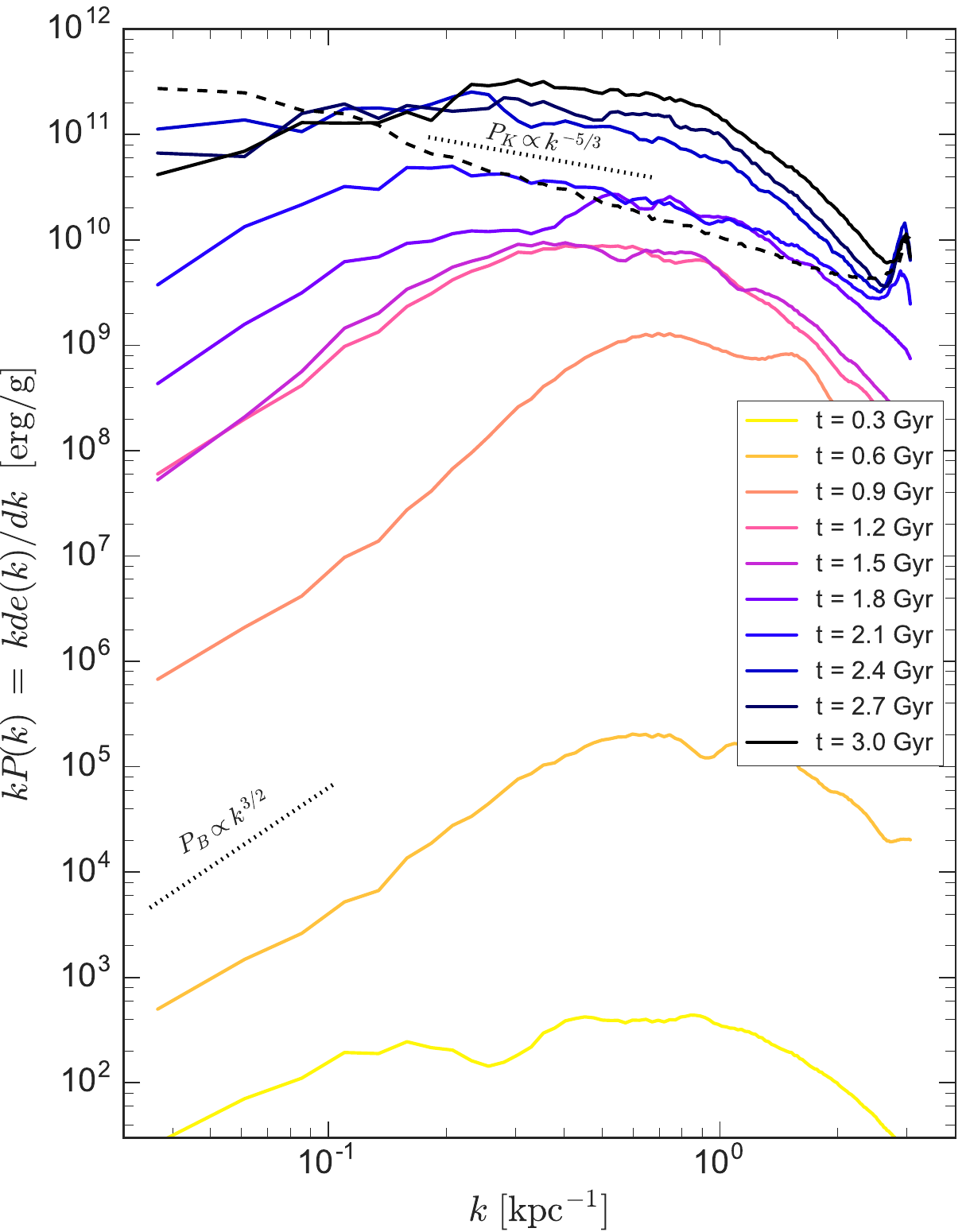}
   \caption{Power spectrum of magnetic energy at listed times in a
     model of an isolated galactic disc.  These spectra show the Kazantsev
     slope at large scales, and a peak that occurs at increasing scale
     over time, as expected from a saturating SSD. The data and the guiding
     lines ({\em dotted lines}) have been regularised by $k$. The kinetic power
     spectrum at the final time is also shown ({\em dashed line}),
     with a slope steeper than expected for incompressible Kolmogorov
     turbulence, consistent with the compressible turbulence in this galaxy. Image reproduced with permission from \cite{butsky2017}, copyright by AAS.}
   \label{fig:kazant-gal-spect}
\end{figure}

The second diagnostic of SSD action is the development of Kazentsev
spectra in models followed by the shift of the peak in magnetic energy
from the dissipation scale to longer wavelengths as saturation sets in
at the dissipation scale, as shown, for example, in
Fig.~\ref{fig:kazant-gal-spect}.  This result has been found in
full-galaxy models by
\cite{rieder2016,rieder2017,rieder2017a,butsky2017,pakmor2017,
  martin-alvarez2018,steinwandel2019,martin-alvarez2022} and
\cite{pfrommer2022}.  \cite{steinwandel2019} and \cite{ntormousi2020}
noted that after saturation, an Iroshnikov spectrum with magnetic
energy proportional to $k^{-3/2}$ appears, while \cite{pfrommer2022}
argues that the saturated spectrum is closer to Kolmogorov with a
$k^{-5/3}$ dependence, although in an accretion-dominated model of a
primordial galaxy.

\citet{martin-alvarez2018,martin-alvarez2022}, and
\citet{ntormousi2020} using AMR and \citet{steinwandel2020} using SPH
explicitly analysed SSD and LSD in global galaxy simulations including
a subgrid model of SN feedback \citep{springel2003a,dubois2008} They
all compared models with and without feedback and found that both
showed the same exponential growth during the first 0.5~Gyr.  At the
resolution of these models, this is the expected timescale for the
action of the SSD, suggesting that the primary driver of the
turbulence required for SSD may be accretion \citep{klessen2010}
rather than SNe in these models that include circumgalactic gas.

\begin{figure}[ht]
\centering
\includegraphics[width=0.9\textwidth]{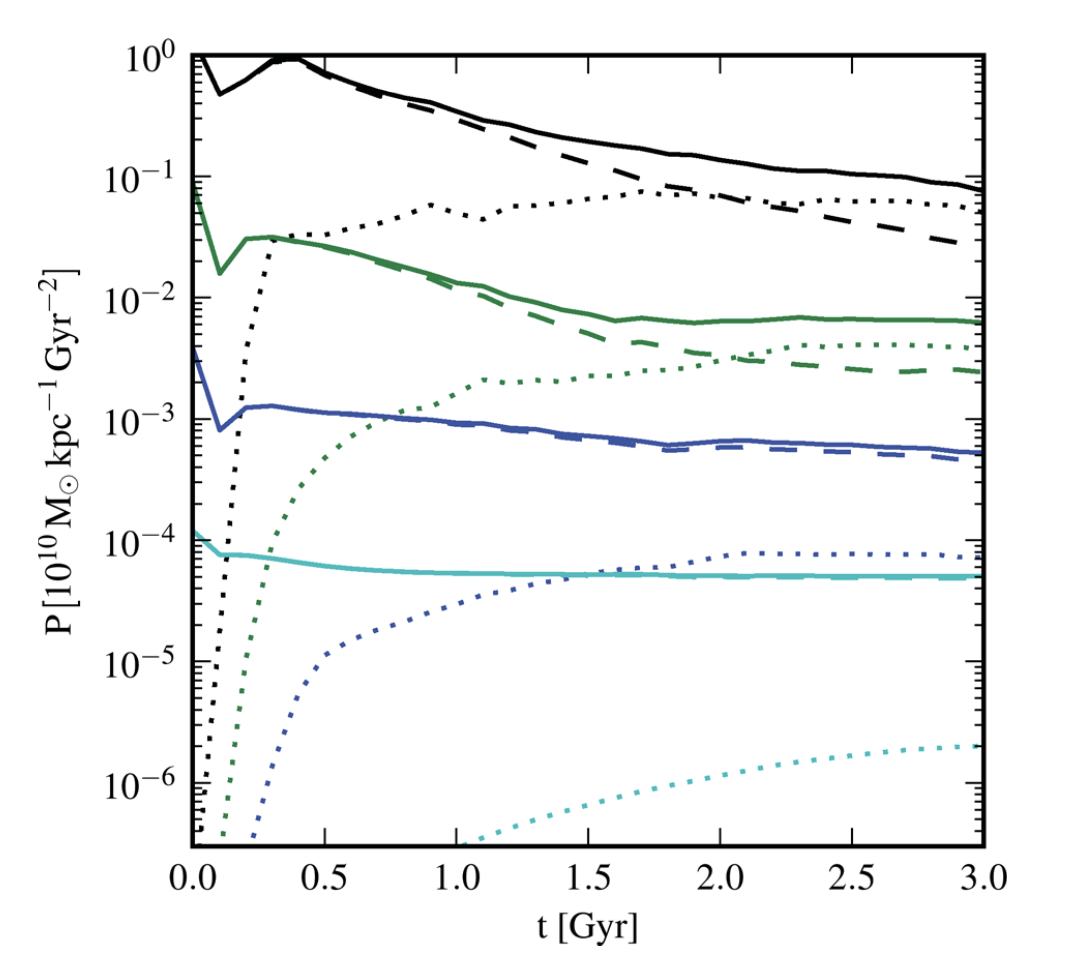}
   \caption{Magnetic (dotted lines), thermal (dashed lines), and total
     (solid lines) pressure for halos with masses of $10^9$ to
     $10^{12}\,M_{\odot}$. All pressures are volume-averaged over the
     galactic discs.  These models show LSD behavior whose
     relative and absolute saturation strength depends strongly on potential well depth. Figure from
     \cite{pakmor2013}.  \label{fig:halos}}
\end{figure}

\paragraph{Large-scale dynamo}
Evidence for the presence of the LSD in global models including SSD has only started
to be presented within the last several years. An early paper that
likely showed LSD action, but did not discuss it explicitly, is
\cite{pakmor2013}.  They compared the growth of volume-averaged
magnetic pressure to that of total and thermal pressure in galaxies of
halo mass varying from $10^9$ to $10^{12}\,M_\odot$, so from dwarf to
Milky Way sizes.  As shown in Fig.~\ref{fig:halos}, all of these
galaxies showed at least modest further growth in magnetic pressure
after saturation of the SSD.  However, the saturation of this apparent
LSD depended strongly on the halo mass, with the smallest, slowest
rotating halos with the shallowest potentials having magnetic pressures
more than an order of magnitude below the thermal pressure, while the
Milky Way mass halo had magnetic pressure exceed thermal pressure.

\begin{figure}[htp]
\centering
  \includegraphics[width=0.6\textwidth]{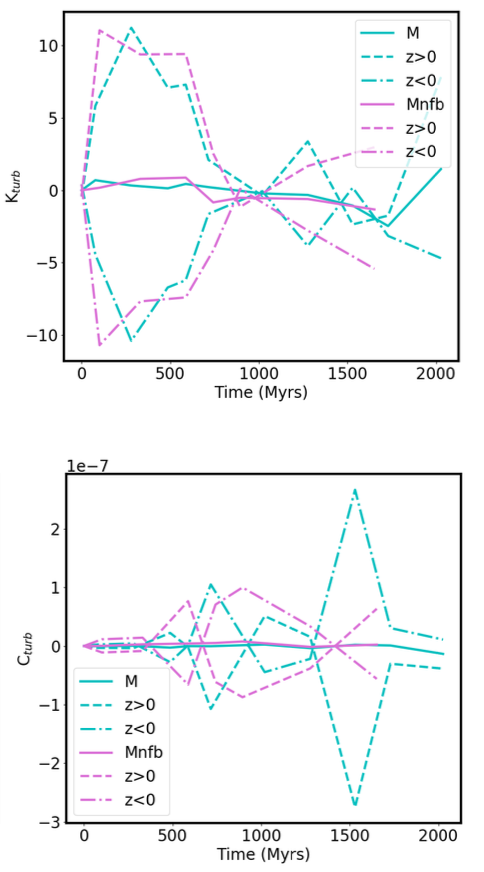}
   \caption{Integrated turbulent kinetic and current helicities over time above
     and below the galactic midplane in a global AMR model embedded in
     a halo. The symmetric sign reversals suggest an unsaturated dynamo. Image reproduced with permission from \cite{ntormousi2020}, copyright by ESO.}
   \label{fig:helicity}
\end{figure}

As helical turbulence is
\mjk{believed to be one of the most important turbulent ingredients}
for an LSD,
\cite{ntormousi2020} directly measured the turbulent kinetic and current
helicities (see Eqs.~\eqref{eq:taua} and \eqref{eq:taub}).
A median filter with 390~pc (8 cell) size was used to separate
turbulent and mean components of the velocity and magnetic field.
In these AMR models, the magnetic energy remained below 0.1\% of the
turbulent kinetic energy, so these are representative of an
unsaturated dynamo.  This is supported by the multiple sign reversals
across the midplane seen in Fig.~\ref{fig:helicity}.

\begin{figure}[ht]
  \centering
      \includegraphics[width=0.49\textwidth]{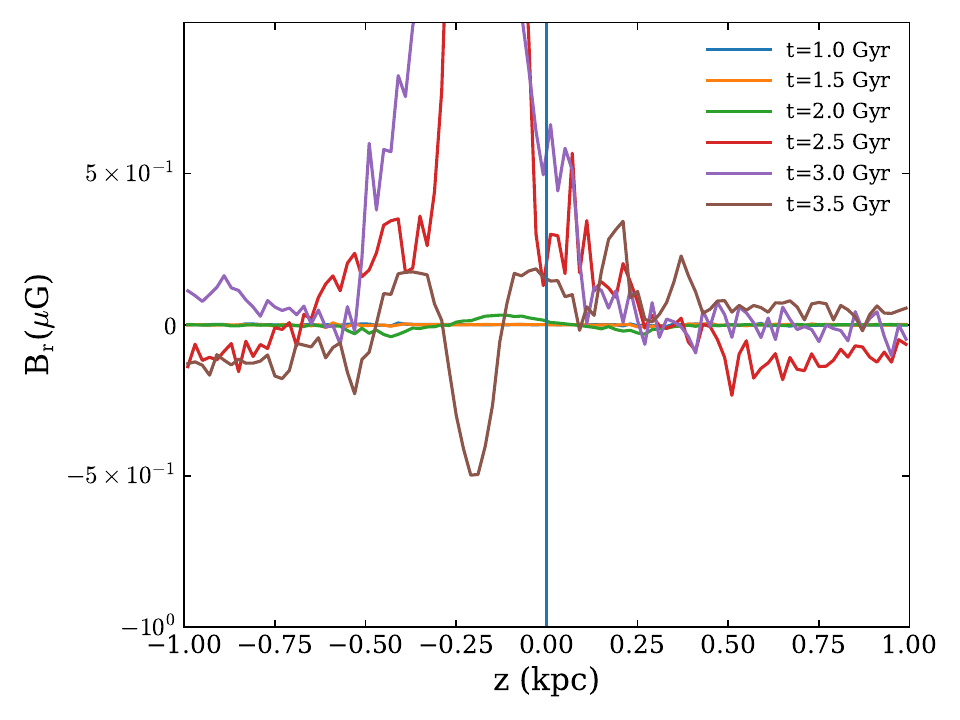}
      \includegraphics[width=0.49\textwidth]{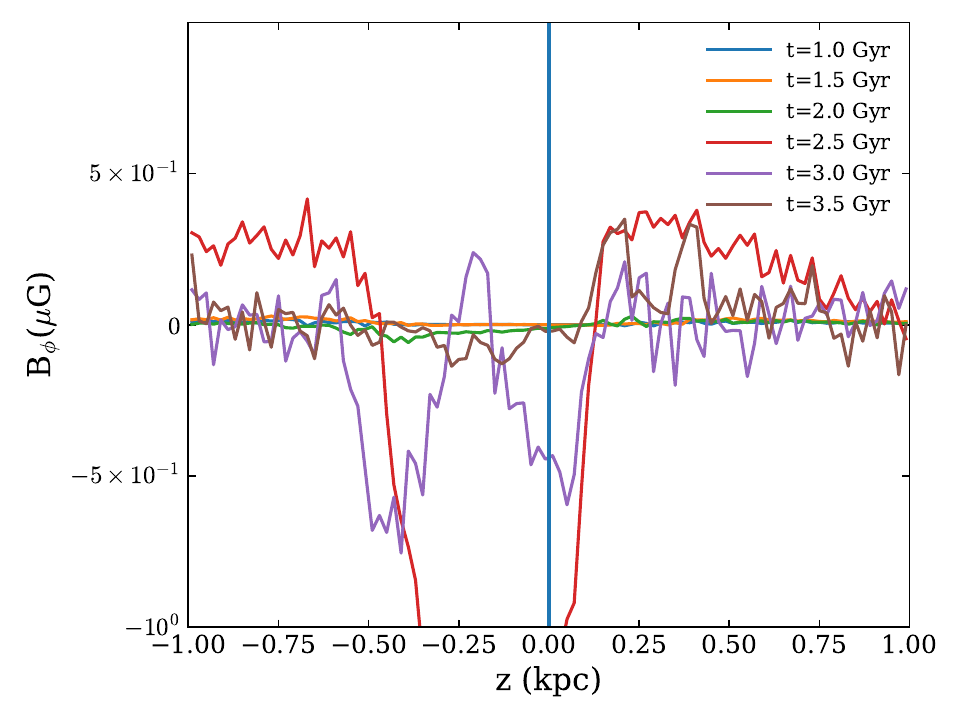}
   \caption{Vertical dependence of radial {\em (left)} and toroidal {\em
       (right)} magnetic fields at times given in the legend for an
     SPH global model. Antisymmetry around the midplane suggests
     dipole structure, while symmetry suggests quadrupolar structure.
     Some indications of both structures are seen at different times,
     but the full structure is more complex, and influenced by the
     buildup of a magnetic tower flow. Image reproduced with permission from
     \cite{steinwandel2020}, copyright by the author(s).}
   \label{fig:polar}
\end{figure}

\cite{steinwandel2020} and \cite{ntormousi2020} both found that the classical
dipolar and quadrupolar field structures do not fully describe the
large-scale field resulting from their models of LSD in global
galaxies, with a rather more confused pattern probably
consisting of multiple eigenfunctions growing simultaneously instead
appearing, as shown for example in Fig.~\ref{fig:polar}.  Similar
results were found by \cite{pakmor2018} and \cite{reissl2023} while
computing Faraday rotation measures, as discussed in
Sect.~\ref{subsec:comparisons} below.

\subsection{Cosmological scale} \label{res:cosmo}

The inclusion of the cosmological environment in models of dynamos in
global galaxies allows a more self-consistent treatment of accretion
on to galaxies and the influence of interactions with neighbors.  The
cost is a low numerical resolution compared to the models of single
galaxies discussed in the previous section.  However, even with SPH mass
resolutions as high as $2\times 10^7\,M_\odot$, \cite{beck2012,beck2013}
demonstrated that an SSD produces exponential field growth during
cosmological halo formation, accretion, and merging even though disc
formation was not resolved.  Radiative cooling balanced by star
formation driven feedback (implemented with a star formation threshold
of only 0.13~cm$^{-3}$ in these low resolution models) produced the
fastest growth rates, with e-folding times of order $10^7$~yr, and
equipartition being acheived with the turbulence.  Saturation occurred
with field strengths of $10^{-5}$~G at the centers of halos.

More recent simulations have followed the formation and evolution of
star-forming galaxies by zooming in to cosmological simulations,
improving the numerical resolution within galaxies in order to resolve
dynamo action there.  \cite{pakmor2014} used the Voronoi-mesh code
AREPO to model a Milky-Way sized galaxy with mass resolution of
$5\times 10^4\,M_\odot$ and a gravitational softening length of
340~pc, while \cite{rieder2017a} used the AMR code RAMSES to model a
dwarf galaxy down to redshift $z = 4$ with most refined resolution of
22.5 physical pc (requiring successive refinements during expansion of
the universe).  These workers found that an SSD drives initial
exponential field growth.  The small-scale turbulence is initially produced by
chaotic accretion, and then subsequently by star formation.  In these
works, increasing the resolution increases the growth rate of the SSD
as expected.  After saturation of the SSD, they find linear growth of
the LSD in the disc, while fields in the galactic center remain at the
value produced by saturation of the SSD absent significant rotation.

Larger scale efforts include the Auriga simulations \citep{grand2017}, a set of
30 zoom-in models using AREPO of Milky Way mass haloes drawn from the dark
matter only version of the Eagle model \citep{schaye2015} at the same mass
resolution of $5\times 10^4\,M_\odot$ used by \cite{pakmor2014}.  Although
these models still do not capture the physical growth rate of the interstellar
SSD  (Sect.~\ref{res:kpc}), they do follow the galactic LSD acting on the
fields generated by halo-scale SSD action during the galaxy formation process.
As the galactic LSD appears insensitive to resolution in this regime
(Sect.~\ref{res:kpc}), the structures modeled in this way may reflect physical
processes and not just numerical dissipation.

\begin{figure}[ht]
\centering
  \includegraphics[width=\textwidth]{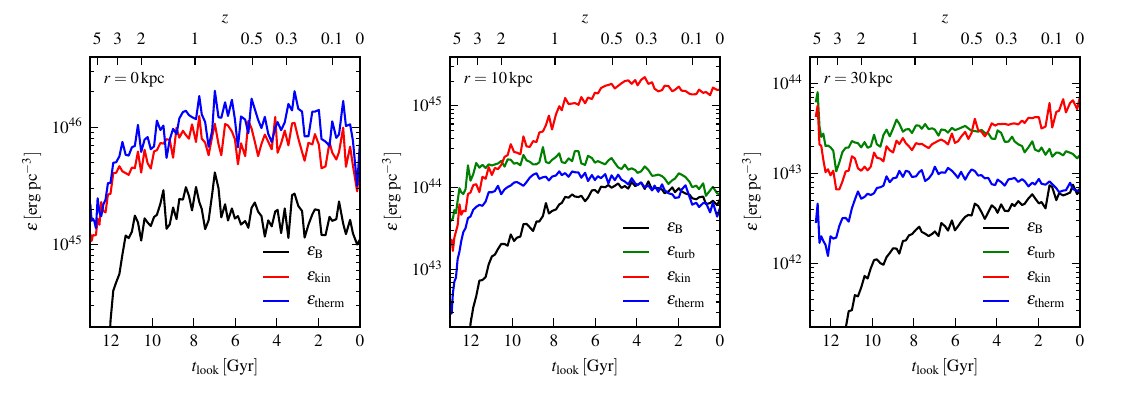}
   \caption{Median energy densities over time drawn from the 30 zoom-in models
     of the cosmological Auriga simulations.  Measurements are
     performed in rings at the given radii.  At 10~kpc galactocentric radius, the
     magnetic field only approaches equipartition with the turbulent
     kinetic energy after 10 Gyr of evolution.  Image reproduced with permission from \cite{pakmor2017}, copyright by the author(s).}
   \label{fig:auriga-median}
\end{figure}

Perhaps the highest resolution cosmological zoom-in model to date is
described by \cite{martin-alvarez2023}, who have simulated a dwarf
galaxy in its cosmological context with 7~pc finest grid resolution
using the RAMSES AMR code.  The MHD runs were terminated at redshift
$z = 3.5$ for reasons of computational cost.  At that time, fields of
2--5~$\mu$G had built up in their discs, consistent with observed
values of $z = 0$ dwarf galaxies \citep{chyzy2011}. Unfortunately, these
models have not yet been analysed in any detail for their dynamo properties.

\subsection{Observational comparisons}
\label{subsec:comparisons}
The correlation between spiral structure in magnetic fields and spiral
structure in stars and gas is observed to be weak
\citep[e.g.][]{Beck15,bittner2017}, with the strongest fields often appearing
in the interarm regions.
This behavior is reproduced by the global model of a
Milky-Way mass galaxy including a circumgalactic medium by
\citet{steinwandel2019}. They
hypothesise that the stronger turbulence in the lower-density, hotter
interarm regions leads to faster field growth there.

\begin{figure}[ht]
\centering
  \includegraphics[width=\textwidth]{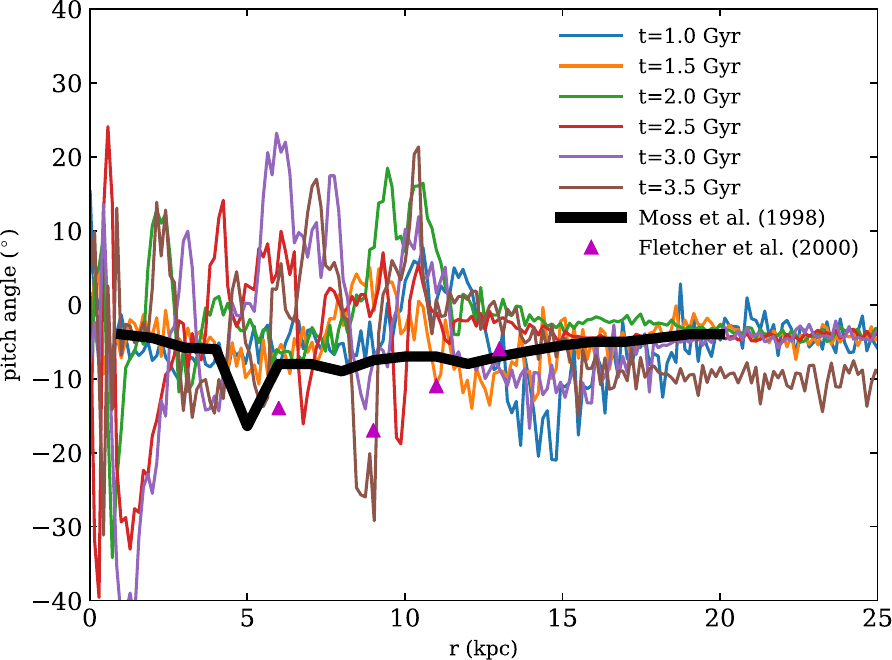}
   \caption{Radial profiles of average pitch angle of magnetic field
     over time for a Milky Way mass galaxy, at times given in the
     legend. Negative pitch angles consistent with observations
     predominate, although positive pitch angles occur in regions
     heavily perturbed by magnetic tower outflows.  The {\em purple
       triangles} show observations of M31 by \cite{fletcher2000};
     these are consistent with the Milky Way value of $-15^{\circ}$ at
     the Solar circle.   The {\em black line} is from a
     dynamo model of M31 by \cite{moss1998}.  Image reproduced with permission from \cite{steinwandel2020}, copyright by the author(s).}
   \label{fig:pitch}
\end{figure}

The pitch angle of the spiral structure in magnetic fields is another
point of comparison between global simulations and observations.
Models show a great deal of scatter in the pitch angle, as shown in
Fig.~\ref{fig:pitch}, with shallow negative pitch angles consistent
with observations being frequent, but by no means universal.  The
situation is further confused by accretion, outflows, and interactions
with other galaxies, all of which can distort the magnetic field
sufficiently for the pitch angle to become positive or strongly
negative.

\begin{figure}[ht]
\centering
  \includegraphics[width=\textwidth]{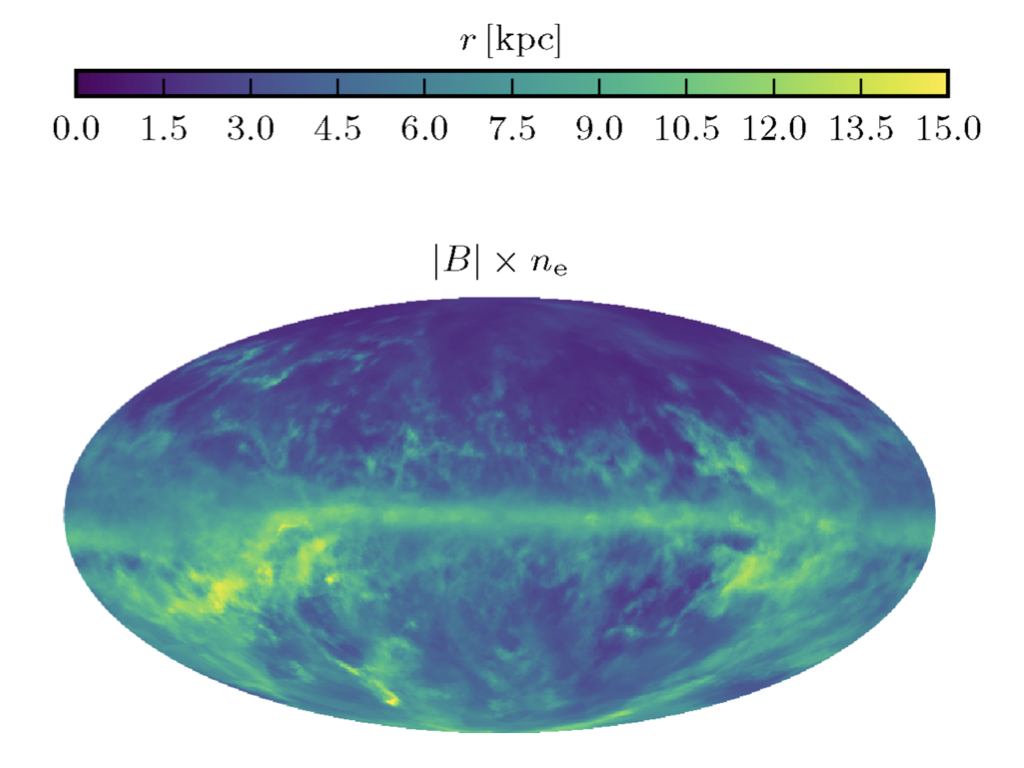}
   \caption{Average depth along the line of sight of contributions to
     the product of the magnitude of the magnetic field and the
     density of thermal electrons measured by rotation measure.  The
     simulated observations are performed for one of the Auriga
     zoom-in models of galaxies in a cosmological simulation.
     Particularly at high latitudes, the dominant contribution comes
     from the local neighborhood rather than the disc as a whole.  Figure adapted from
     \cite{pakmor2018}.  \label{fig:Bn-depth}}
\end{figure}

\begin{figure}[ht]
\centering
  \includegraphics[width=\textwidth]{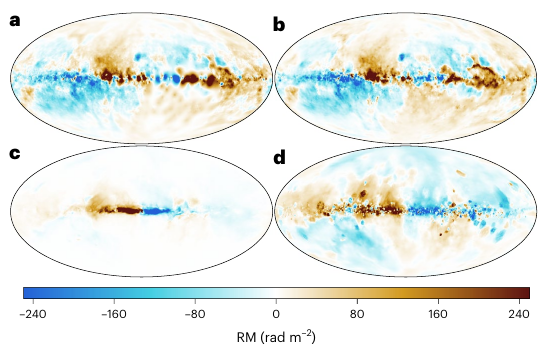}
   \caption{Comparison between the all-sky rotation measure maps
     derived by {\em (a)} \cite{opperman2012} and {\em (b)}
     \cite{hutschenreuter2022} to the rotation measure map from {\em
       (c)} an
     unmodified Auriga zoom-in model of a galaxy, and {\em(d)} from the same model
     and viewpoint but with a subgrid model of ionising radiation
     from star clusters determining the thermal electron distribution.  The
     additional small-scale structure reproduces the statistical small-scale
     structure distribution of the observations far better. Image reproduced with permission from \cite{reissl2023}, copyright by the author(s).}
   \label{fig:RM}
\end{figure}

\citet{carteret2023} used a shearing box model of a section
  0.8~kpc on a side of a
galactic disk to measure the structure of Faraday rotation and
polarisation due to dynamo-derived fields.  They found that different
assumptions about the
thermal electron density distribution, not explicitly modeled in their
simulation absent radiative transfer of ionising radiation, drive
leading order variation in the magnitude and scale distribution of the
rotation measure for the same magnetic field structures.  Anisotropic
structure in the rotation map occurred under both assumptions, though.

\cite{pakmor2018} and \cite{reissl2023} modeled observed Faraday
rotation on the sky as seen from within one of the galaxies from the
Auriga cosmological zoom-in simulations \citep{grand2017} at roughly
the Solar circle. These models were compared to the observed
distributions derived by \cite{opperman2012} and
\cite{hutschenreuter2022} from measurements towards extragalactic
sources.  \cite{pakmor2018} found agreement on the strength and
qualitative distribution of intermediate to large scale structure in the distribution,
but found much less small-scale structure than in the
observations. They found that the structure within a few kiloparsecs
dominates the Faraday rotation, particularly at high latitudes, rather than any global dynamo-formed
dipole or quadrupole field, as shown in Fig.~\ref{fig:Bn-depth}. \citet{reissl2023} used cluster population synthesis and
explicit transfer of ionising radiation \citep{pellegrini2020} to
model the distribution of thermal electrons down to a much smaller scale, allowing polarised
radiative transfer calculations from a vantage point within a
superbubble at roughly the Solar radius. Consistent with the results
of \citet{carteret2023}, they find that the addition
of the subgrid model of cluster ionisation dramatically improves the
agreement of the model with the observations at smaller scales as
shown in Fig.~\ref{fig:RM} and quantified in the paper.


\begin{figure}[ht]
\centering
\includegraphics[width=0.9\textwidth]{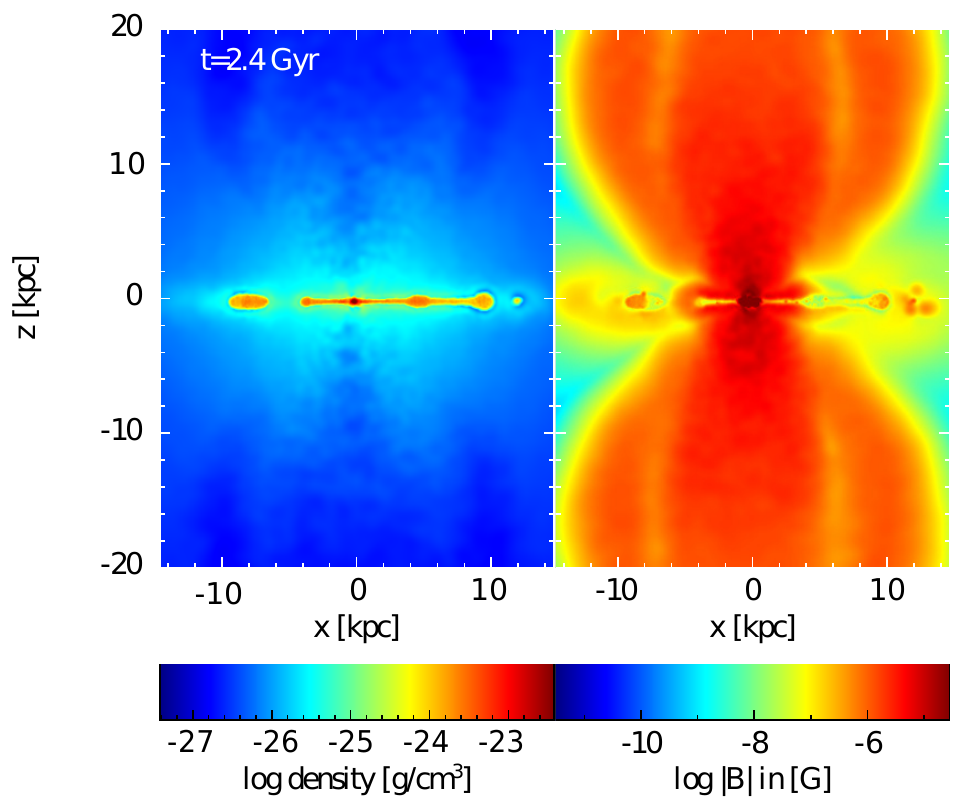}
   \caption{Slices of density and magnetic field magnitude for
     Milky-Way mass models including circumgalactic medium structure
     after 2.4 Gyr of evolution. The magnetic tower outflow driven by magnetic
     pressure produces X-shaped structures in the circumgalactic
     medium similar to those observed. Image reproduced with permission from
     \cite{steinwandel2019}, copyright by the author(s).}
   \label{fig:tower}
\end{figure}

A strong magnetic tower flow (Fig.~\ref{fig:tower})
with a velocity of a 400--500~km~s$^{-1}$ driven by the
dynamo-driven increase of magnetic
pressure in the center of the modeled galaxy occurs in the model of
\cite{steinwandel2019}.  This structure appears consistent with the X-shaped magnetic
field structures seen emerging from disc galaxies in observations of
radio polarisation \citep[e.g.][]{veilleux2005}.
However, this is a
transient effect that lasts only a few hundred megayears in the model,
raising the question of how such a structure can be commonly observed
in modern galaxies.

\begin{figure}[ht]
\centering
  \includegraphics[width=0.9\textwidth]{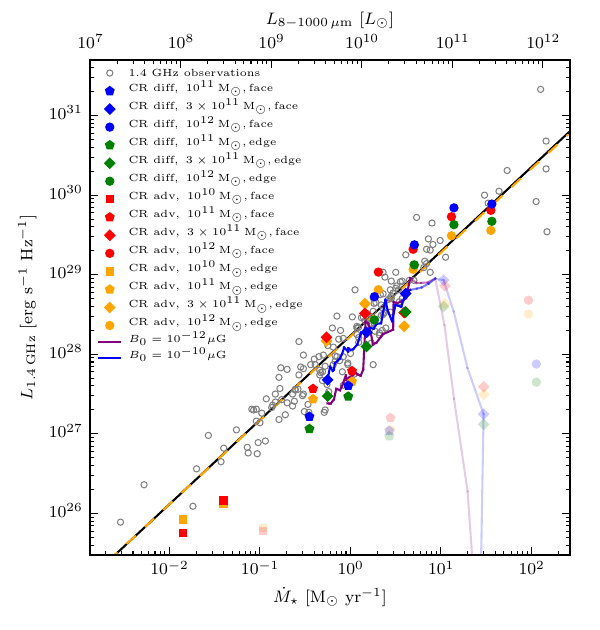}
   \caption{Comparison of observations of the correlation between
     radio synchrotron luminosity at 1.4 GHz and far infrared
     luminosity at 8--1000 $\mu$m {\em (open circles)} to luminosities
     computed from simulations of
     accretion dominated galaxies with the Voronoi mesh code AREPO. The
     semi-transparent symbols and tracks show results from unsaturated
     dynamos and their evolution, while the solid colors represent
     fully saturated dynamos. Models with only cosmic ray advection
     are labeled ``CR adv'' in the legend, while models also modelling
     cosmic ray diffusion are labeled ``CR diff''.   The legend also
     gives halo masses and orientation for the different galaxies. Image reproduced with permission from \cite{pfrommer2022}, copyright by the author(s).}
   \label{fig:synchFIR}
\end{figure}

A detailed comparison of the interaction between fields generated
by the SSD and cosmic rays in an accretion dominated galaxy to the
correlation between
far-infrared and radio synchrotron luminosities was presented by
\cite{pfrommer2022}. Fig.~\ref{fig:synchFIR} shows that the
correlation can be reproduced by fully saturated SSDs in these
galaxies, a result that is robust to the details of the modelling of
cosmic ray travel and the orientation of the galaxies.
\citet{carteret2023} examined polarisation of synchrotron emission in
their shearing box model with an assumed distribution of cosmic rays.
They showed that the cosmic ray distribution and the field both matter
for the synchrotron emission, and that the observed structure further
depends on the wavelength of observation.

Another comparison between simulations including cosmic rays and
observables was done by \cite{ponnada2022}.  They compared synthetic
rotation measures to observational constraints from the circumgalactic
medium around galaxies.  They found rotation measures consistent with
observed upper limits, but low enough that they would not be detected
with current instruments, suggesting that improved observations will
be needed to fully constrain these models.

\section{Properties of turbulence}

Numerical models have been used to extract dynamo-relevant properties
of turbulence. Such studies have two purposes. Firstly, comparisons
with analytical theories can be made. Secondly, some ill-known
parameters and properties can be deduced, which can be vital to
constructing simpler and less resource-demanding numerical approaches.

\begin{figure}[ht]
  \includegraphics[width=\textwidth]{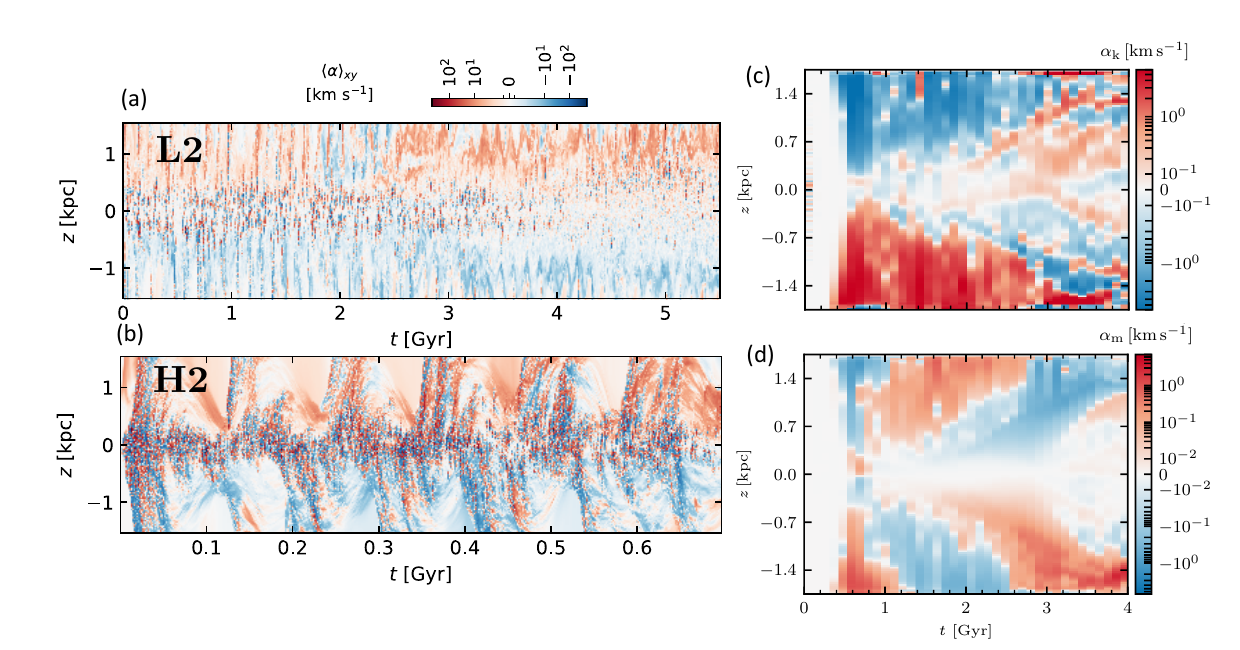}
  \caption{Time-latitude diagrams of horizontally averaged
    $\alpha$-effect for a SN-driven model with twice the galactic
    rotation and shear rate from {\em (a)} a lower-resolution model
    integrated up to the saturation of the dynamo and {\em (b)}
    a higher-resolution model with a far shorter time extent, but
    illustrating the detailed time evolution. Adapted from
    \cite{Gent23SSDLSD}. On the right we show  {\em (c)} the kinetic $\alpha$
    value and {\em (d) } the magnetic $\alpha$ value from a
    Parker-unstable system without SNe, but with galactic rotation
    rate. Adapted from \cite{Devika_Parker23}. \label{FredDevika} }
  \end{figure}

\subsection{Retrieval of turbulent transport coefficients} \label{TFM:results}

Semi-analytic approaches \citep[see, e.g.][and references
  therein]{Ferriere98} showed early on that isolated SNe cause only
weak turbulent effects relevant for the dynamo, while superbubbles
could produce more significant effects in the required range for
producing a dynamo number above the critical one needed for LSD
action. These models, however, excluded SN interactions, such as
colliding shock fronts, where non-linear interactions, such as
generation of vorticity through the baroclinic effect (arising when
temperature and entropy gradients are misaligned) or vortex
stretching, have been reported to generate significant amounts of
vorticity \citep[see, e.g.,][]{,Korpi:1999c,Vort2018} To study such
effects, full MHD models are needed.

The most straightforward way of estimating the turbulent inductive and
diffusive effects from numerical simulations is to use kinetic and
magnetic helicity and turbulent intensity as a proxy for them (see
Eqs.~\eqref{eq:taua} and \eqref{eq:taub}).
To date, this method has only rarely been implemented
to diagnose the full MHD simulations.
Recently,
\cite{Gent23SSDLSD} computed the proxy for the $\alpha$ effect this
way in their SN-forced, stratified, rotating and shearing local model,
exciting both SSD and LSD, and ran their lower-resolution models until
the saturation of the LSD (see Fig.~\ref{FredDevika} {\em (a)} and {\em (b)}). In
the growth stage of the LSD the kinetic $\alpha$ dominates, is
smooth and of the expected sign ($+$ in the upper and $-$ in the lower part of the
disc) in regions of more laminar inflows caused by cooling. On the
other hand, it is
strong and spatially incoherent with undetermined sign in the strong
outflows caused by clustered SN activity. The unexpected sign of
$\alpha$ in the latter case seems to originate from the opposite disc plane, which
indicates that some non-trivial helicity fluxes might occur
across the midplane of the disc. The magnetic $\alpha$ appears
to have two clear roles: to revert the sign of the $\alpha$
effect during a reversal of the
azimuthal  magnetic field
at approximately 2~Gyr, and to saturate LSD growth in the warm gas
concentrated around the midplane.

\cite{Devika_Parker23} derived the $\alpha$ contributions in the same
way from a Parker unstable stratified disc otherwise similar, but
without SNe and Milky Way solar neighborhood rotation and shear rates
((Fig.~\ref{FredDevika} {\em (c)} and {\em (d)}), while
\cite{Gent23SSDLSD} had twice the Solar neighborhood values. A
noteworthy feature, in comparison to the SN-driven system in panels
{\em (a)} and {\em (b)} is that the kinetic $\alpha$ effect is
systematically of the opposite sign (up to 2 Gyr). The smooth parts of
the $\alpha$ effect in the different cases are roughly of the same
magnitude, while significantly stronger only in the SN bursts and the
resulting outflows. The higher rotation and shear rates in
\cite{Gent23SSDLSD} most likely also enhance the SN-driven $\alpha$
magnitudes, so they might, in fact, be rather equal in a consistently
similar system. The early change in sign of the magnetic part of the
$\alpha$ effect is similarly relatable to a reversal of the azimuthal
magnetic field in the Parker-unstable case. Also, in this case the LSD
is saturated by the growing magnetic contribution of the opposite
sign. Comparing these two studies it is evident that the
Parker-instability induced dynamo is of significance and operates in a
markedly different way than an SN-driven dynamo, while both are
expected to occur in the SN-active part of galactic discs. Hence,
future numerical studies including both these mechanisms is of great
importance. Such studies, however, require large enough domains
  also in the SN-forced models to capture the Parker unstable magnetic
  loops.

\begin{figure}[ht]
  \includegraphics[width=\textwidth]{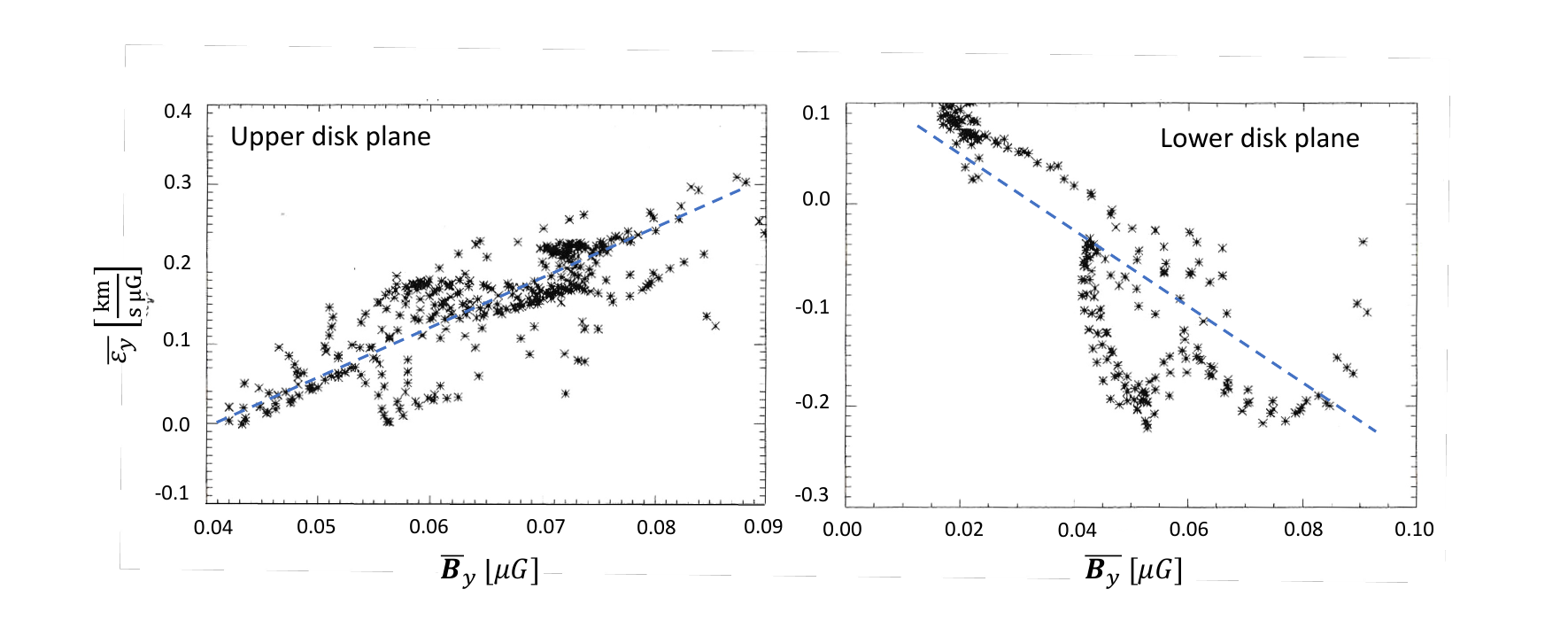}
  \caption{The mean azimuthal magnetic field vs.\ the mean
    electromotive force. Averages are taken over space and over a 10
    Myr time interval, separately for the {\em (left)} upper and {\em (right)} lower
    disc plane of the Cartesian shearing box simulation. This
    time period roughly corresponds to the turnover time in the local
    simulation domain. The blue dashed lines show $\alpha_y = \pm 6$
    km s$-1$ correlation coefficient from the relation
    $\overline{\emf}_y=\alpha_y \Bm$. Adapted from
    \cite{Korpithesis99}.}\label{Korpithesis99}
  \end{figure}

Another numerical method to retrieve a limited set of the turbulent
transport coefficients is the imposed field method. It relies on the
Taylor expansion of the mean emf according to Eq.~\eqref{emf} under
the assumption that all higher order terms than the linear term
containing the $\alpha$ coefficients vanish; this requires the
  imposed magnetic field to be uniform. One imposes a field in some
direction, and then measures the mean electromotive force components
and the mean magnetic field, and finally solves for the relevant
$\alpha$ tensor component(s). The problem is that non-uniform magnetic
fields can be induced, leading to the formation of currents relatively
quickly.  These currents cause the breakdown of the method unless the
magnetic field is reset at suitable time intervals \citep[see,
  e.g.,][]{OSBR02,Hub+09}. This method has not been applied very extensively
in the context of galactic dynamos, although early attempts
(Fig.~\ref{Korpithesis99}) resulted in a successful
measurement of the main $\alpha$ effect component generating the mean
radial field from the azimuthal one \citep{Korpithesis99}. There was
no explicit imposed field applied, as the system itself was dynamo
active, and resulted in the amplification of mean fields. Due to
poor resolution, the dynamo was near its critical Re$_{\mathrm{M}}$ for
being active. Therefore, the field never reached high strength due to the
limited duration of the simulation runs, and thus remained in the
kinematic regime. Indications of an $\alpha \Omega$ dynamo with
an $\alpha_y$ coefficient of roughly $\pm 6\,\mathrm{km\ s}^{-1}$ were found. This
was a factor of a few in excess of the nearly contemporary
semi-analytic findings that neglected SN interactions, thus indicating
that the SNe are, indeed, important.

\cite{BNST95} studied accretion disc dynamos driven by MRI in a
Keplerian shearing box model. They found a self-sustained dynamo, and
applied a similar method to measure the $\alpha_y$ component as
\cite{Korpithesis99}. As a result, a clear correlation between the
mean emf and theazimuthal magnetic field was again found, but the sign of
$\alpha_y$ was the opposite: negative in the upper and positive in the
lower hemisphere, similar to the Parker-unstable system of
\cite{Devika_Parker23}. This suggests that dynamos driven by magnetic
buoyancy might all have this feature in common. Also galactic discs
are susceptible to
MRI \citep{piontek2005, piontek2007}
although their rotation laws
are not Keplerian, but typically flat. Therefore, although the \cite{BNST95} results are not directly applicable,
they could indicate that multiple different dynamo mechanisms could be at
play in galaxies, if SN-driven and MRI-driven turbulence
co-exist. This, however, seems unlikely, as SN activity has
analytically been shown to suppress MRI in the inner 15 kpc region in
a Milky Way like galaxy \cite{KKV10}. MRI-driven turbulence and
dynamos could, however, explain the unexpectedly large velocity
dispersions in the extended HI discs of galaxies
\cite{SB99a,tamburro2009}.

A method suitable for numerically extracting the turbulent transport
coefficients from anisotropic systems, such as galactic discs, which
definitely have strong anisotropies arising from stratification,
(non-uniform) rotation, and the presence of large-scale magnetic
fields, was developed by \cite{BS02}. In such systems, $\alpha$
and $\beta$ are tensors instead of scalars. Thus, to
extract all the tensor components, one needs many
equations. \cite{BS02} proposed that enough equations can
be collected by forming various moments of the mean field components
and their derivatives. Next, correlations with turbulent
  emf, again utilising the ansatz Eq.~(\ref{emf}), are sought for.
All of these
quantities are measurable from a full MHD
simulation.
Their formulation of the resulting multidimensional
regression problem was either spatially local, considering the value
of the magnetic quantities and the turbulent emf at one single point
only, or else non-local, considering a convolution kernel around the
location. This method is referred to as the method of moments or the
correlation method, and it has been extended to further improve
the fitting procedure by using SVD
\citep[][]{SCD16,Bendre2020,Bendre2022}, which usually prevents
unphysical solutions such as negative values of $\beta$. The method is known to
work best for time-variable magnetic fields, and to underestimate the
magnitude of $\beta$ especially for fields stable over time
\citep[][]{Warnecke2017a}.

\begin{figure}[ht]
  \includegraphics[width=\textwidth]{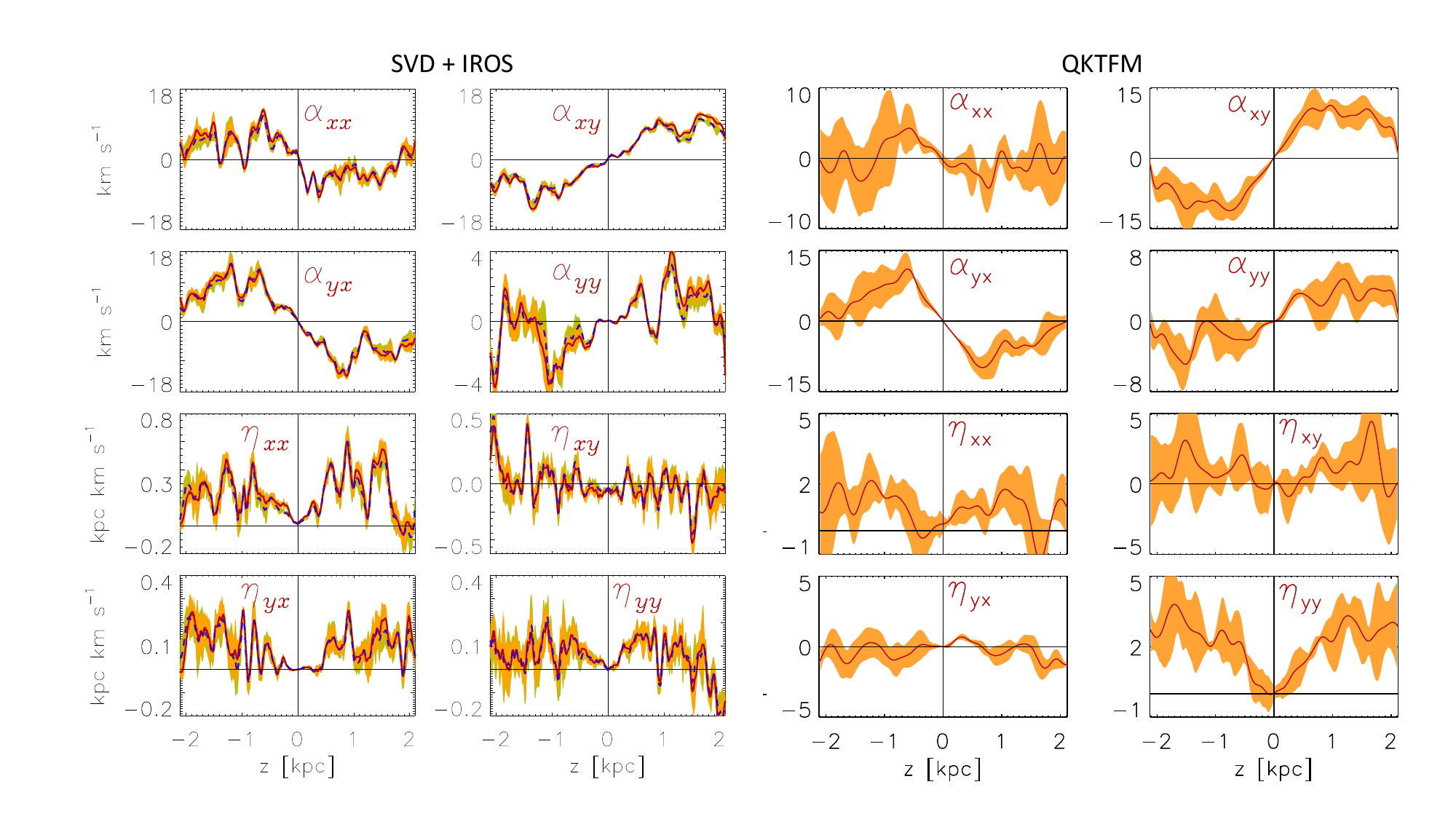}
  \caption{Derived turbulent transport coefficients from three
    different methods: {\em (left)} local SVD {\em (red, errors in
      orange)} and IROS {\em (blue dashed, errors in green)} results
    from \citet{Bendre2023}; {\em (right)} QKTFM results from
    \cite{Bendre2020}. The DNS-like MHD model
    from which these coefficients are derived is Model Q in
    \citet{BGE15}, a local, Cartesian, SN-driven simulation with
    $\Omega = 4\OSN$ and $\sigma = 0.25 \ZSN$. The model develops
    large-scale dynamo action with oscillatory behavior in the
    kinematic stage, and dynamically significant, but stable in time,
    magnetic fields in the non-linear stage. The turbulent transport
    coefficients are derived from the kinematic
    stage.  \label{Bendre}}
  \end{figure}

An illustration of local SVD results from a local Cartesian
MHD simulation with the NIRVANA code are shown in the leftmost set of
Fig.~\ref{Bendre}, where components of the $\alpha$
tensor
and the most relevant components of the
turbulent resistivity tensor are shown as a function of the
vertical height from the midplane. The overall sign convention and
peak values of the $\alpha_{yy}$ component are rather similar to those found by
\cite{Korpithesis99}, although the profiles are quite complex and
there are even sign reversals as functions of height, and strong
gradients near the boundaries. The profiles of the diagonal turbulent resistivity
tensor components are also complex, and unexpected negative values occur
throughout the domain, especially near the top of it.

Recently, \cite{Bendre2023} developed a novel method, called iterative
removal of sources (IROS), for extracting the turbulent transport
coefficients. This method shares many similarities with the SVD
method, while the main differences are the iterative fitting
algorithm, which incrementally refines the estimates of the turbulent
transport coefficients, and the possibility to define prior physical
constraints (mIROS scheme). The latter property significantly
decreases the errors, and the processing of extremely noisy data
becomes possible. The first results of the IROS method are illustrated
in Fig.~\ref{Bendre} leftmost set, and compared to the SVD method. As
is evident, these two methods give nearly identical results.

An important approach to
measuring the turbulent transport
coefficients is
the TFM (\citealt{Schrinner05}, \citealt{Schrinner07}, recently reviewed in
detail by
\citealt{AxelMFreview18}).  In short, this method is based on the
exertion
of linearly independent test fields
to the system that are passive in the sense of not
interfering with it, but act as a diagnostic tool. The action of the
turbulent flow on the test fields is monitored, and the equation for
the fluctuating fields is directly solved for each test field.
Turbulent emf is approximated as local and instantaneous and its
  expansion is truncated after first order of the derivatives of
    $\Bm$(Eq.~\eqref{emf}).  This approximation is again used to
solve for the turbulent transport coefficients. By directly
solving for the fluctuating fields, many detrimental assumptions of
the FOSA/SOCA are avoided. The local and instantaneous formulation
imposes the restriction of capturing only effects of large-scale,
slowly varying in time MFs. This restriction can, however, be relaxed
by considering a formulation non-local in space and time \citep[see,
  e.g.][]{RB2012}. Typical results from QKTFM in the kinematic stage
of a large-scale, dynamo-active, MHD model are shown in
Fig.~\ref{Bendre}, rightmost set. Similar kinematic analysis on MHD
models with varying parameters has been performed by
\cite{gressel2008,Gressel08a}, and \citet{BGE15}.

In comparison to SVD, the profiles from QKTFM are overall somewhat
smoother, while the correspondence of the magnitudes and vertical
profiles of the $\alpha$ coefficients between
the methods is satisfactory. The most notable difference is in the magnitude of the
turbulent resistivity components. Both diagonal elements are roughly
an order of magnitude larger with the QKTFM than with SVD. The
off-diagonal component $\eta_{xy}$ is consistent with zero from SVD,
while it has a clearly positive value from the
QKTFM. On the other hand, $\eta_{yx}$ is consistently non-zero and
positive from SVD, while consistent with zero from QKTFM. This does rule
out the existence of the shear-current dynamo effect in galactic discs by
both methods, the existence of which would require $\eta_{yx}$ being
negative\citep[see, e.g.,][]{IgorNathan03}.

The quenching of the turbulent transport coefficients in the non-linear
regime, due to the presence of strong large-scale magnetic field, has
also been studied with the QKTFM by \cite{Gressel2013b}. Finally,
\cite{Gressel2020} studied the importance of non-local and
non-instantaneous effects in similar models, and found out that
especially the former are important for galactic LSD. The $\alpha$ and
$\eta$ coefficients were found to reduce when the scale of the
test fields is made smaller, agreeing with results from simpler
systems \citep[e.g.][]{BRS08}. A surprising finding was that the
turbulent pumping, generally directed towards the disc plane and
aiding LSD, was becoming stronger with decreasing scale. This behavior
is yet to be fully understood.

Until recently, DNS-like models to study
galactic LSD have hovered at resolutions too coarse to properly
capture SSD simultaneously. Such models are now emerging \citep{Gent23SSDLSD}. In these
models dynamically significant LSD is excited despite the
co-existence of fast-growing SSD.
This might not be so surprising, as an SSD alone in a
multiphase, highly compressible fluid is weak, producing fluctuating
fields with an energy of only a few percent of the kinetic energy in the
turbulence \citep{gent2023}. Also, these systems might exhibit
much larger helicity fluxes than the simpler systems studied
previously (see Sect.~\ref{HelicityFluxes:results}).

In this situation,
determining the turbulent transport coefficients with the QKTFM is no
longer straightforwardly possible, as the SSD generates
background magnetic fluctuations
that are not accounted for in the method. Viable
extensions \citep{RB10,SMHD,CTFM} have
appeared, but even further extensions, such as relaxing isothermality, are
required to deal with the galactic models. Furthermore, only the
kinematic version of the compressible test-field method could be
validated for very strong magnetic fields \citep{CTFM}. Also, the
required computational resources for the extraction of the turbulent
transport coefficients significantly exceeds that for
QKTFM. Evidently, SVD and IROS methods would not suffer from any of
the discussed limitations, although the better noise-handling capacity
of mIROS may turn out to be necessary.

Although the results of the different methods seem to be in rough
agreement, it is still necessary to ask the question of which of the
methods is the most reliable? The fitting-based methods SVD and IROS seem to agree
well with each other, but they are both based on the use
of the MF of the MHD simulation itself in the fitting
procedure. In contrast, the test-field-based methods
use a set of linearly
independent and well defined test fields. In the
fitting approach, the
MF generated by the system could be degenerate, while the
test-field approach guarantees that such a situation cannot occur, and the
coefficients can be unambiguously determined. Also, the TFMs are
rigorously tested against analytical solutions such as various
incarnations of the Roberts flow \citep[see, e.g.,][]{RB10}, while
such verifications have not been performed on the fitting-based
methods. In \cite{Gressel2020} the authors propose that the
scale-dependence of the transport coefficients could provide an
explanation to the discrepancies between SVD and QKTFM, but this
appears unlikely.
Namely, the SVD approach uses the
DNS-generated MF scale for the determination of both the $\alpha$ and $\eta$
coefficients, which both would become reduced, if the proposed explanation would be valid.
Only the $\eta$
coefficients, however, show marked reduction.

The
best way to test
these discrepancies would be to compare the evolution of MF models using the measured
coefficients to the actual MHD
models. Curiously, this has been done with the SVD
coefficients by \cite{Bendre2020}, showing the two solutions to be in
excellent agreement in the oscillatory kinematic regime. The period
of dynamo cycles is known to
directly depend on the turbulent resistivity
\citep[see, e.g.,][]{SS22}.
Then one would expect to see a discrepancy between the
QKTFM and MF results. This remains to be confirmed, and
conclusions drawn accordingly.

The ultimate goal of the efforts to measure and describe the
turbulent dynamo coefficients relevant for the LSD in galaxies is to
build computationally more efficient global modelling frameworks. In
essence, the goal is to build explicit large eddy simulations, where also the backscatter
(or inverse cascade from small scales to larger ones) is captured in as
much detail as possible. This has been attempted either in the MF
framework \citep{Gressel13} or in a simpler context for
  cosmological modelling
\citep{liu2022}. The most promising explicit large eddy frameworks exist only for
simpler turbulence models \citep[see,
  e.g.,][]{grete2017}.

\subsection{Measuring helicity fluxes} \label{HelicityFluxes:results}

Helicity fluxes have also been measured from various types of MHD
models. Such work is vital, as the final proof that helicity fluxes
actually alleviate catastrophic quenching must come from full MHD
models. The measurements necessarily involve many simulations with
varying $\Rm$, reaching out to as high $\Rm$ as
possible. This is demanding, so models remain limited to simplified,
isotropically
forced systems.

The rate of change of
the mean helicity density of the small-scale field reads
\begin{equation}
  \frac{\partial}{\partial t} \overline{{\bf a'} \cdot \Bf} = -2 \emf
  \cdot \Bm - 2 \eta \mu_0 \overline{{\bf j'} \cdot \Bf} - \nabla
  \cdot \overline{{\bf F}}, \label{helevol}
\end{equation}
where the first term on the right hand side describes the generation of magnetic
helicity and the second one its resistive destruction,
while $\overline{\bf F}$ is the small-scale helicity flux \citep[see,
e.g.,][]{BS05}. If the flux is significant, its divergence should be
dominant over the two other terms during a steady state dynamo. From the
evolution equation for the dynamical $\alpha$ effect,
Eq.~\eqref{eq:alpdyn},
one would also expect that the mean field strength
would become independent of $\Rm$, if the fluxes are efficient.

\cite{DelSordo13} studied an isothermal plane-wave forced Cartesian
system with a possibility for an imposed wind in one coordinate
direction. The forcing used was helical, producing an equator
perpendicular to the wind direction by changing the sign of
  helicity across the equatorial plane. Although simplified, this
model has some similarities to a local volume of a rotating galactic
disc subject to a wind. The results show that the wind-like advection
can induce a non-zero magnetic helicity flux, which shows asymptotic
behavior as a function of $\Rm$. The other terms in Eq.\eqref{helevol}
were larger for $\Rm < 1000$, but for higher $\Rm$ they became
comparable to the divergence of helicity flux. Moreover, these terms
showed
an inverse dependence on
$Rm$. However, the MF decreased when the wind strength
increased. Essentially similar results were obtained with a comparable
but incompressible setup without a wind \cite{Rincon21}. Agreeing with
the results of \cite{DelSordo13} without the wind,
\cite{Rincon21} found significant helicity fluxes due to the
redistribution of helicity through turbulent flux across the equator.
Although promising, these results do not yet
conclusively prove that helicity fluxes can alleviate the catastrophic
quenching scenario. Simulations with higher $\Rm$ and also including
shear and more realistic turbulence driving are required.

 \section{Conclusions}


 \subsection{Comparisons with analytic theory}

 Many aspects of analytic theory appear consistent with numerical
 simulations, in particular the existence of both SSD and LSD activity in galaxy
 models at different scales. SSD growth rates are strongly dependent
 on $\Re$ (Eq.~\ref{eq:Gamma}), and thus on numerical resolution if it dominates
 resistivity over any physical resistivity implemented.  SSD and LSD
 compete for energy when both are active, but after SSD saturates, LSD
 continues to grow MF as predicted analytically (Sect.~\ref{kpc:lsd}).  The requirement of
 helical turbulence for dynamo activity is verified (Sect.~\ref{subsub:gal-mhd}).
 Parker instability, MBI, and MRI can drive further dynamo activity,
 although likely subdominant effects in star-forming regions of
 galactic disks (Sect.~\ref{kpc:lsd}.

 Turbulent transport coefficients derived from the numerical models with
 multiple different sets of assumptions show that, although the
 overall picture of dynamo behaviour from analytic models does have
 some resemblance to the results, there are many inconsistencies (Sect.~\ref{TFM:results}).
 These are most likely due to the strong, anisotropic turbulent flows
 produced by the SN driving in stratified discs.  As a result, simple
 dipolar or quadrupolar field configurations are likely to be more the exception
 than the rule (Sect.~\ref{subsub:gal-mhd}).

 The question of the source of the magnetic fields that grow due to
 dynamo activity has been dramatically answered in recent years with
 the clear demonstration of dynamo action leading to
 near-equipartition fields occurring in plasmas
 magnetised {\em ab initio} by Weibel instability (Sect.~\ref{res:plasma}). As this can occur already in
 protogalactic accretion flows, galaxies appear likely to be born
 magnetised.  Fast SSD action in these high Re systems maintains
 fields within an order of magnitude of equipartition on less than megayear
 timescales.  Large-scale MF does take much longer to develop, as it
 still relies on LSD activity over far longer timescales comparable to
 the rotational periods of galaxies  (Sect.~\ref{res:kpc}).

 Many models suggest that helicity fluxes can prevent catastrophic
   quenching of LSD activity by small-scale magnetic fluctuations
   (Sect.~\ref{HelicityFluxes:results}).  However, direct measurements
   of helicity fluxes in well-defined numerical experiments will still
   be required to conclusively demonstrate that catastrophic quenching
   does not occur in galactic dynamos.

\subsection{Consistency with observations}
The observed global structure of magnetic fields has points of contact
with models.  The weak correlation of magnetic spiral arms with stars
and gas is seen in models including a circumgalactic medium.  Models
show a wide variety of pitch angles, some of which do appear
consistent with measurements of shallow negative values, although even
positive values are seen in the models in perturbed or strongly
interacting galaxies (Sect.~\ref{subsec:comparisons}).

Simulated maps of Faraday rotation derived from models agree best with
observations if whole galaxy models are combined with a sub-grid model
for H~{\sc ii} regions produced by stellar clusters drawn from a
population synthesis model.  High-galactic latitude regions in
particular are dominated by structure within a few kiloparsecs of the
Sun, if not the surface of the Local Bubble.  Predictions of rotation
measures are consistent with observed upper limits, but small enough
that improved observations will be needed to constrain them (Sect.~\ref{subsec:comparisons}).

\subsection{Challenges}

Starting from the earliest times, models of seed field production and
amplification by Weibel instability need to be extended towards
realistic values of the electron-ion mass ratio, and checked for
robustness against Landau damping.  Quantitative demonstration of
their applicability to protogalactic accretion flows will also be
valuable.

However, the strong implication of these models that galaxies are born
magnetised suggests that numerical models of galaxy formation and
evolution should not start with infinitesimal
fields, but rather with turbulent fields of some fraction of
equipartition, and that those minimum values of the turbulent field
should be somehow maintained even if the numerical resolution of
global or cosmological models is insufficient to directly capture the
relevant fast SSD maintaining the models.
Performing simulated observations of the synchrotron radiation and IR
polarisation produced by the MF structure of galaxies in
their cosmological context will ultimately require models of both
magnetic field and cosmic ray evolution.  These will allow much better
constraints on the models to be placed by classical observations such
as the pitch angle of spiral arms and field reversals.

Achieving sufficient resolution to measure MF dynamo coefficients in
models that include the necessary physical processes in the ISM is a
third challenge, that offers a potential solution to the second
challenge.  If dynamo coefficients can be accurately and
self-consistently measured, they can be used to support the
construction of explicit large eddy simulations that capture the
sub-grid dynamo behavior.  Although preliminary work indicates
agreement between MF models in some regimes and the DNS-like MHD
models from which they draw their coefficients
(Sect.~\ref{TFM:results}), much further work needs to be done to
extend this result into saturated regimes.  Disentangling the effects
of multiple types of dynamos acting simultaneously remains a major
challenge.

\backmatter

\bmhead{Acknowledgments}
We acknowledge fruitful discussions with Abhijit Bendre, Matthias
Rheinhardt, and J\"orn Warnecke on the manuscript.  This project has received funding from the European Research Council
(ERC) under the European Union's Horizon 2020 research and innovation
program (Project UniSDyn, grant agreement n:o 818665; for MJKL and
FAG). M-MML was partly supported by US NSF grant AST23-07950 and
acknowledges Interstellar Institute's program ``II6'' and the
Paris-Saclay University's Institut Pascal for hosting discussions that
nourished the development of some of the ideas behind this work.

\paragraph*{Disclosure statement}

The authors declare that they have no relevant or material financial
interests that relate to the research described in this paper.

\phantomsection
\addcontentsline{toc}{section}{References}
\bibliography{mara,maclow,fred}

\end{document}